\documentclass[fleqn,usenatbib]{mnras}

\usepackage{newtxtext,newtxmath}

\usepackage[T1]{fontenc}
\usepackage[normalem]{ulem}

\DeclareRobustCommand{\VAN}[3]{#2}
\let\VANthebibliography\thebibliography
\def\thebibliography{\DeclareRobustCommand{\VAN}[3]{##3}\VANthebibliography}

\usepackage{longtable}
\usepackage{natbib}
\usepackage{color}
\newcommand{\project}[1]{\textsl{#1}}

\newcommand{\apogee}{\project{\textsc{apogee}}}
\newcommand{\lamost}{\project{\textsc{lamost}}}

\newcommand{\Gaia}{\project{Gaia}}
\newcommand{\Gaiaeso}{\project{Gaia--\textsc{eso}}}
\newcommand{\galah}{\project{\textsc{galah}}}

\newcommand{\rgal}{\mbox{$\rm R_{gal}$}}

\usepackage{multirow}
\usepackage{multicol}

\usepackage{graphicx}	
\usepackage{amsmath}

\title[GalStar]{Milky Way Mapper decoded abundances -- II: From patterns to paths}

\author[M. K. Ness et al.]{Melissa K. Ness$^{1}$\thanks{E-mail: melissa.ness@anu.edu.au},
Sarah Aquilina$^{1}$,
Jennifer Mead$^{2}$,
Emily Griffith$^{3}$,
Catherine Manea$^{4}$,
Jonathan Bird$^{5}$,
\newauthor
Andrew R. Casey$^{6,7}$,
Lucy (Yuxi) Lu$^{8,9}$,
Kathryn V. Johnston$^{2}$,
Michael R. Blanton$^{10}$,
James W. Johnson$^{11}$,
\newauthor
Maja Jablonska$^{1}$,
Leticia Carigi$^{12}$,
Jos\'e G. Fern\'andez-Trincado$^{13}$,
Ricardo L\'opez Valdivia$^{14}$,
Ying-Yi Song$^{15,16}$,
\newauthor
Juna Kollmeier$^{11,17,18}$, 
\\
$^{1}$Research School of Astronomy and Astrophysics, Australian National University, Canberra, ACT 2611, Australia\\
$^{2}$Department of Astronomy, Columbia University, Pupin Physics Laboratories, New York, NY 10027, USA\\
$^{3}$Center for Astrophysics and Space Astronomy, Department of Astrophysical and Planetary Sciences, University of Colorado, 389 UCB, Boulder, CO 80309-0389, USA\\
$^{4}$Department of Physics \& Astronomy, University of Utah, Salt Lake City, UT 84112, USA\\
$^{5}$Department of Physics and Astronomy, Vanderbilt University, 6301 Stevenson Center, Nashville, TN 37235, USA\\
$^{6}$Center for Computational Astrophysics, Flatiron Institute, 162 5th Ave, New York, NY 10010, USA\\
$^{7}$School of Physics and Astronomy, Monash University, VIC 3800, Australia\\
$^{8}$Department of Astronomy, The Ohio State University, 140 W 18th Ave, Columbus, OH 43210, USA\\
$^{9}$Center for Cosmology and Astroparticle Physics (CCAPP), The Ohio State University, 191 W. Woodruff Ave., Columbus, OH 43210, USA\\
$^{10}$Center for Cosmology and Particle Physics, Department of Physics, New York University, 726 Broadway Rm. 1005, New York, NY 10003, USA \\
$^{11}$Carnegie Science Observatories, 813 Santa Barbara Street, Pasadena, CA 91101, USA\\
$^{12}$Universidad Nacional Aut\'onoma de M'exico. Instituto de Astronom\'ia. A.P. 70-264, 04510. Ciudad de M\'exico, M\'exico \\
$^{13}$Universidad Cat\'olica del Norte, N\'ucleo UCN en Arqueolog\'ia Gal\'actica - Instituto de Astronom\'ia, Av. Angamos 0610, Antofagasta, Chile\\
$^{14}$Universidad Nacional Aut\'onoma de M\'exico. Instituto de Astronom\'ia. A.P. 106, 22800. Ensenada, B.C. , M\'exico \\
$^{15}${David A. Dunlap Department of Astronomy \& Astrophysics, University of Toronto, 50 St. George Street, Toronto, ON M5S 3H4, Canada} \\
$^{16}${Dunlap Institute for Astronomy \& Astrophysics, University of Toronto, 50 St. George Street, Toronto, ON M5S 3H4, Canada} \\
$^{17}$Canadian Institute for Theoretical Astrophysics, University of Toronto, Toronto, ON M5S 3H8, Canada\\
$^{18}$Canadian Institute for Advanced Research, 661 University Avenue, Suite 505, Toronto, ON M5G 1M1, Canada 
}

\date{Accepted YYYY Month DD. Received YYYY Month DD}

\pubyear{2024}

\begin{document}
\label{firstpage}
\pagerange{\pageref{firstpage}--\pageref{lastpage}}
\maketitle

\begin{abstract}
The element abundances of Milky Way disc stars encode entangled imprints of multiple enrichment processes, making it difficult to uncover the underlying chemical evolution. Here we re-project  16 stellar abundances for $\approx 199,290$ red giant stars ([Fe/H]$ > -1$) into a set of (4) shared enrichment patterns, providing a generative framework for learning the organising structure of the Milky Way disc. The relative contributions of these patterns vary systematically across the disc, revealing a low-dimensional enrichment basis that responds coherently to global drivers of disc evolution. By grouping stars according to their pattern contributions, we identify coherent enrichment pathways that exhibit strong chemo-spatial correlations and are stratified in both age and height above the plane, linking radial growth to vertical disc structure. Stars occupying similar positions along these enrichment pathways also show coherent vertical deviations across radius, indicating that the low-dimensional chemical structure captures the disc’s response to dynamical perturbations. We identify a transition in enrichment behaviour at approximately 6~Gyr, marking the onset of a more chemically mixed regime with increasing contributions from delayed sources. Within this connected system, the observed $\alpha$-bimodality arises within a shared, low-dimensional abundance structure, with stars populating continuous sequences of changing enrichment fractions that are tightly coupled to spatial, temporal, and orbital coordinates across the Milky Way disc. 
\end{abstract}


\section{Introduction}

Stellar abundances trace birth environment conditions and contain information about the gas cloud in which they formed. In the Milky Way disc these are correlated with, but not directly set by, a star's present-day spatial and orbital properties. Processes including radial migration and dynamical heating redistribute stars away from their birth sites over time \citep{Selwood2002, Roskar2008, Minchev2011, RS2009}. However, chemo-dynamical correlations reveal that stellar populations retain a memory of their original spatial conditions \citep[][]{BH2010, Bovy2016b, Minchev2018, Ratcliffe2021, Ness2019, Mackereth2017, Rix2013, SandersDas2018}. Observations show that stars with similar orbital actions or ages often share distinctive abundance patterns, reflecting their common origin despite subsequent migration. \citep{Bovy2016b}

Large-scale spectroscopic surveys including \apogee, \lamost, \galah\ and \Gaiaeso\ have revealed that the Milky Way disc formed inside-out, with radial migration playing a significant role in shaping its present-day structure \citep{Frankel2018, Frankel2019}. The \textit{Gaia} mission has been transformative, providing precise parallaxes as well as velocity measurements that show the Galactic disc is not in dynamical equilibrium \citep[e.g.,][]{Hunt2025}. Clear signatures of phase mixing and disequilibrium have been observed \citep{Antoja2018, Belokurovsplash, Hawkins2023, Wheeler2022a, Orti2023}. Combining detailed elemental abundances from spectroscopic surveys with kinematics and orbital information from \textit{Gaia} has supported extensive chemo-dynamic studies of the disc including connections to inferred birth radius \citep{Minchev2018, Wang2024, Ratcliffe2021, Lu2024}. The combination of dynamics, spatial information, abundances as well as stellar age estimates, has been key in studying the temporal evolution of the disc \citep[e.g.,][]{Mackereth2017, Bonaca2020, Belokurov2022, Chandra2023, Ness2016, Rix2013, Casali2023,  Horta2024, Bensby2003, Ratcliffe2024, Bovy2016b, Adi2013, Bovy2012, Eilers2022, Ratcliffe2023, Anders2023, Minchev2017, Xiang2022, Ness2019}.

In physical space, the Milky Way disc is characterised by a divide into a thin and thick component \citep{Gilmore1983, Fuhrmann1998}. James Webb Space Telescope (JWST) observations show that disc galaxies similar to the Milky Way are in place at high redshift \citep{Kuhn2024}. Notably, the emergence of a distinct thin disc appears to have occurred in a mass-dependent fashion: more massive galaxies developed thin discs earlier in cosmic time than less massive galaxies \citep{Takafumi2025}.  These findings suggest that the formation of Milky Way-like discs is a common and relatively continuous process across cosmic time. In the Milky Way the thick disc comprises older, more $\alpha$-enhanced stars with larger velocity dispersions and greater scale heights, while the thin disc is made up of younger, less $\alpha$-enhanced stars that remain closer to the Galactic mid-plane \citep{Hayden2015, Nidever2014, Queiroz2020, Imig2023, Horta_oti, Chandra2023, SilvaAguirre2018, Bensby2003}. 

The origin of the thin and thick disc and corresponding $\alpha$-bimodality has been a subject of substantial interest across observational, theoretical, and simulation studies. Broadly, the distinct structural properties and chemical gradients of the disc components are thought to reflect, though with varying emphasis across studies, the combined effects of early formation history, merger events, quenching episodes, and internal processes such as radial migration and disc heating. The $\alpha$-abundance bimodality and its differences in stellar age and structure has been interpreted as a signature of two-phase formation and disc settling over time. In this scenario, an early, rapid burst of star formation produces the high-$\alpha$ population. This is followed by a stall or quenching phase, and then a more prolonged period of star formation that builds the low-$\alpha$ sequence \citep[e.g.,][]{Lian2020a}. This can be modeled in a `closed box' framework, where there is a fixed and available gas reservoir \citep[e.g.,][]{Haywood2013, Snaith2015}. This framework emphasises the consistency with radial abundance trends, whereby the high-$\alpha$ sequence is dominant in the inner Galaxy and the low-$\alpha$ in the outer \citep{Hayden2015,  Nidever2014, Queiroz2023}. Other models use an `open box' framework with either discrete or continuous gas inflows, with the gas supplied by the  circumgalactic medium or merger-sourced \citep[e.g.,][]{Chiappini1997, Spitoni2023, Lian2020a, Brook2012, Buck2020, Kh2021, Spitoni2021, Orkney2025, Johnson2025}.  An alternative (but not necessarily mutually exclusive) explanation is that the thick disc has arisen from a continuous early disc that was dynamically heated by mergers or internal perturbations \citep[e.g.,][]{Villa2010, Grand2020}, while star formation continued in a thinner disc afterward. Both frameworks require a delay between the early high-$\alpha$ and later low-$\alpha$ star formation epochs to account for the observed chemical discontinuity.

Several observational and theoretical studies consider the Milky Way’s disc as a single, continuous system rather than as two entirely separate components. In this view, the thick and thin disc or high and low-$\alpha$ populations emerge naturally from the continuum of star formation and enrichment, without requiring a clean break between them \citep{Schonrich2009, Haywood2006, Haywood2013, Sharma2021, Chen2023, Bovy2012, Bovy2012b, Mackereth2017, Minchev2017}.  In such models, the disc is understood as an organised, evolving ensemble of stellar populations with age-dependent scale heights and gradually changing chemistry;  a continuum of ``mono-age'' or ``mono-abundance'' populations. The apparent gap in the abundance plane then reflects a transitional period in the Galaxy’s star formation and enrichment history, namely when Type~Ia supernovae began to substantially enrich the interstellar medium, not necessarily accompanied by a stop and start of disc formation. The chemo-dynamical simulation of \citet{Kh2021} shows that the  $\alpha$-bimodality and other disc phenomena can be understood as a consequence of continuous disc growth modulated by dynamics, without invoking distinct mergers and subsequent gas infall episodes. 

Internal dynamical processes associated with the Galactic bar may also play a significant role in the disc’s chemo-structural evolution. The bar’s resonances can redistribute stars and gas, potentially contributing to features in the age-metallicity relation or the $\alpha$-bimodality that would not appear in axisymmetric models. For instance, a slowing (decelerating) bar can induce radial migration by ``dragging'' stars outward from the inner Galaxy at specific resonances. Such bar-driven migration mechanisms have been proposed as an important factor in disc evolution, including the formation of the alpha-bimodality  \citep[][]{Beane2018, Zhang2025}. Other studies focus on the role of external perturbations in terms of gas enrichment changes \citep[e.g.][]{Lu2024, RL2020, Agertz2021, Buck2020}.

If the Milky Way disc formed in a continuum process, stellar abundances should show a \textit{shared} chemical evolution pathway between thin and thick discs and, correspondingly, low and high-$\alpha$ populations. There is already evidence that the high-$\alpha$ and low-$\alpha$ sequences share chemical evolution pathways. The low-dimensional structure of the disc’s full abundance space means that both high- and low-$\alpha$ stars can be described using a compact set of common nucleosynthetic sources.  Data-driven models trained using only a few abundances, or sources, across the entire disc (without separating thick/thin) are able to predict more than a dozen, and up to 30, individual stellar abundances with high accuracy \citep{Ness2022, Weinberg2019, Griffith2024, Mead2025}. This indicates that all disc stars, whether in the high-$\alpha$ or low-$\alpha$ class, share a largely coherent chemical enrichment history.  At the same time, the existence of the $\alpha$-bimodality means that a \textit{transition} in the dominant enrichment pathways did occur. 

In this work, we examine the chemo-dynamical state of the disc using the data-driven framework of the Non-negative Matrix Factorisation (NMF) of abundance patterns as described in \citet{Ness2026} (hereafter Paper I). Under this approach, the vector of stellar abundances of the Milky Way disc stars are re-projected onto a set of shared enrichment patterns. In this follow up work we analyse a larger set of stars across the disc, tracking the inter and intra group behaviour of stars organised by their fractional pattern similarity.  In Section~\ref{data} we summarise the data used in our analysis and revisit the mathematical formalism of the latent basis. In Section~3 we show that the spatial behaviour of the four latent channels is governed by a shared radial basis, and we group stars by their fractional contributions to examine the intra- and inter-group behaviour. This reveals coherent enrichment signatures that connect the latent chemical channels to the disc’s formation across time and spatial extent. In Section~\ref{discussion} we discuss the implications of these results,
and we conclude in Section~\ref{conclusion}.

\section{Data and Methods}
\label{data}

We use the version-5 of the SDSS-V \textsc{ASTRA} results (Casey et al., in preparation) for the Milky Way Mapper Milky Way Mapper survey \citep[DR19;][]{Kollmeier2025, DR19}. SDSS-V utilises the Sloan Foundation Telescope at Apache Point Observatory \citep{Gunn2006} and the du Pont Telescope at Las Campanas Observatory \citep{Bowen1973}, both equipped with APOGEE spectrographs \citep{Wilson2019}. 

Paper~I established the latent abundance framework using a sample of 70,057 Galactic Genesis red giant stars. In the present work, we apply and extend this framework to a substantially larger data set of 199,290 stars, enabling a more comprehensive characterisation of the spatial and dynamical behaviour of the latent enrichment channels. We verify that the findings here are also in agreement with this analysis performed using only the high-fidelity subset analysed in Paper~I.

To increase the sample size, we revise the stellar selection criteria
from Paper~I by broadening the ranges in effective temperature and
surface gravity, increasing the allowable per-element abundance
uncertainties, and lowering the signal-to-noise threshold. We have validated that very similar results are obtained with our smaller, high fidelity Paper~I sample, indicating that this less stringent restriction does not compromise our results. 

Below we
describe the resulting data set in detail and summarise the method from Paper~I used to factorise the abundances into shared patterns and
per–star coefficients of these patterns.

\subsection{Data}

We select Milky Way Mapper red giant stars from the version-5 of the SDSS-V \textsc{ASTRA} results using the following set of criteria;

\begin{itemize}
  \item Signal-to-noise ratio: $\mathrm{S/N} > 50$
  \item Metallicity: $[\mathrm{Fe/H}] > -1$
  \item Effective temperature: $4500\,\mathrm{K} < T_\mathrm{eff} < 5300\,\mathrm{K}$, with uncertainty $\sigma_{T_\mathrm{eff}} < 100\,\mathrm{K}$
  \item Surface gravity: $0 < \log g < 3.5$, with uncertainty $\sigma_{\log g} < 0.3$
  \item Elemental abundances: all $ -2.3 < [X/\mathrm{H}] < 1$~dex
  \item No flag set `$X\_h\_flags$' for any element abundance
  \item Elemental abundance uncertainties: all $\sigma_{[X/\mathrm{H}]} < 0.20$~dex
  \item Finite abundance and uncertainty values for all elements
\end{itemize}

This gives a selection of 218,449 red giant stars with 16 measured element abundances, elements,  X=Fe,~O,~Mg,~Al,~Si,~S,~K,~Ca,~Ti,~V,~Cr,~Mn,~Fe,~Co,~Ni,~Ce,~Nd,  which are reported as [X/H]. We removed duplicates, and 199,290 remain. Of these stars, $97$\% are within $|z| < 2$~kpc of the mid-plane of the disc and 85\% are within $|z| < 1$~kpc of the mid-plane of the disc.

Similarly to Paper I, we calculate a number of orbital parameters; the angular momentum $L_z$ , the radial action, $J_r$ and vertical action $J_z$, and the eccentricity,  using the \texttt{galpy} package \citep{galpy}, assuming the solar motion from \citet{Schonrich2010}. This is done by combining observed positions and proper motions from \Gaia\ DR3 \citep{GC1}, with radial velocities from Milky Way Mapper, converting them into \texttt{galpy} \texttt{Orbit} objects, and integrating them in the \texttt{MWPotential2014} Milky Way potential model. Adopted distances are the median geometric distance `$r\_{med}\_{geo}$' \citep{BailerJones2021}. We adopt the ages from \citet{StoneM2025} to analyse the results as a function of stellar age. Again, in this work, the orbital parameters serve primarily as descriptive diagnostics rather than inputs to a detailed dynamical model. We exclude $\approx$ 1000 stars from the analysis that have distance uncertainties and proper motion uncertainties greater than 100\%. While the sample ranges from \rgal $\sim 0-20$kpc, approximately 75\% of the sample are within \rgal$\approx 6-10$~kpc.

\subsection{Non Negative Matrix Factorisation (NMF)}
\label{method}

Similarly to Paper~I, we use non-negative matrix factorisation (NMF) to model
stellar abundances as combinations of a small number of shared latent patterns.
We begin with an abundance matrix $X$ of $N$ stars and $K$ elements, and construct
a shifted matrix $X'$ that enforces non-negativity prior to factorisation. We
factorise $X'$ into a basis of $M=4$ shared patterns with per-star coefficients,
using a variance-weighted implementation and the same set of 16 elements as in
Paper~I. In this work we apply the model to a larger sample of
$\approx 218{,}400$ stars.  Following Paper~I, our model is: 

\begin{equation}
    X' \approx f\,P ,
\end{equation}
where $f$ is an $N \times M$ matrix of non-negative pattern coefficients,  and
$P$ is an $M \times K$ matrix of non-negative abundance patterns.

Equivalently, each element of the shifted abundance matrix can be written as
\begin{equation}
    X'_{ij} \approx \sum_{m=1}^{M} f_{im}\,P_{mj}.
\end{equation}

The entries $f_{i,m}$ represent
the coefficients, or weights, of latent pattern $m$ contributing to the
abundance vector of star $i$. Each row of $f$ therefore expresses an individual
star as a non-negative mixture of $M$ latent enrichment channels. Each row $P_m$ corresponds to the characteristic elemental
abundance pattern associated with latent enrichment channel $m$. Under this
model, each stellar abundance vector is represented as a linear combination of
the rows of $P$, weighted by the corresponding coefficients in $f$.

We perform the NMF decomposition on the shifted abundance matrix $X'$ by
minimising a weighted least-squares loss function that accounts for measurement
uncertainties:
\begin{equation}
\min_{f,\,P \ge 0}
\sum_{i=1}^{N} \sum_{j=1}^{K}
W_{ij}\,\bigl(X'_{ij} - (fP)_{ij}\bigr)^2 .
\end{equation}

The weight matrix $W$ has dimensions $N \times K$ and is defined as
\begin{equation}
W_{ij} = \frac{1}{\sigma_{ij}^2},
\end{equation}
where $\sigma_{ij}$ is the measurement uncertainty of element $j$ for star $i$.
We assume the uncertainty covariance matrix is diagonal and neglect
inter-element covariances.

The shifted abundance matrix $X'$ is defined as
\begin{equation}
X'_{ij} = X_{ij} - \min_{j'}(X_{ij'}) + \epsilon ,
\label{eqn:shift}
\end{equation}
where the minimum is taken over all elements $j'$ measured for star $i$, and
$\epsilon$ is a small positive constant added for numerical stability. This
deterministic, per-star offset satisfies the non-negativity requirement of NMF
while removing the overall abundance scale, thereby isolating the intrinsic
pattern structure of the elemental abundances.

To evaluate the goodness of fit, we reconstruct the abundances by multiplying
the inferred matrices $f$ and $P$. All reconstructions are transformed back to
the original measurement scale by re-applying the per-star abundance offset
before computing goodness-of-fit statistics. The per-star $\chi^2$ is then
computed using the same weighted form as the loss function minimised during
factorisation:

\begin{equation}
\chi^2_i =
\sum_{j=1}^{K} W_{ij}
\left[
X_{ij}
-
\left(
(fP)_{ij}
+ \min_{j'}(X_{ij'})
- \epsilon
\right)
\right]^2.
\label{eqn:chi}
\end{equation}

The recovered patterns for our basis of $M=4$ are in agreement with those identified in Paper~I, but now inferred using a substantially larger set of stars across the disc.  Also similarly to Paper~I, the  M=4 latent patterns well capture the 16 disc abundances, as evaluated using the $\chi^2$ (Equation \ref{eqn:chi}). The  mode of the distribution of the reduced $\chi^2$ for the $\approx 199,290$ stars is reduced $\chi_{mode}^2 \sim1.2$, with 75\% (90\%) of stars having a reduced $\chi^2 < 3$ (5).  Therefore, the majority of stars are well represented by the four–latent-variable model, in agreement with the results of Paper~I.  In Paper~I we associated the latent channels with broad nucleosynthetic sources based on their characteristic elemental signatures. The patterns recovered exhibit some (not unexpected) variations relative to the earlier analysis, but remain aligned with the prior assignment. The pattern vectors are shown in the Appendix Figure \ref{fig:appendix1}. We term each of these patterns channels which we associate with signatures of different nucleosynthetic sources.

\begin{itemize}

\item Channel~1 (high O, Mg, Al, Si, S, K, Ca, Ti) represents an ``early'' $\alpha$-pattern, from core collapse supernovae (SN~II) \citep{Nomoto2013}. 
\item Channel~2 (high Mg, Al, K, Ca, Ti) represents a ``late'' $\alpha$-pattern, from core collapse supernovae (SN~II) \citep{Woosley1995, Nomoto2006, CK2020}. 
\item Channel~3 (high Mn, Cr Fe, Co, Ni) is characterised by iron-peak elements, from Supernovae Ia (SN~Ia) \citep{Nomoto1997, CK2020}.
\item Channel~4 (high Ce) shows an s-process signature, from Asymptotic Giant Branch (AGB) stars \citep{Karakas2014}.

\end{itemize}

The validation showing the NMF generated and corrected back to their original reference frame, compared to measured \textsc{ASTRA ASPCAP} abundances are shown in the Appendix in Figure~\ref{fig:appendix2}.

\begin{figure*}
    \centering
    \includegraphics[width=\textwidth]{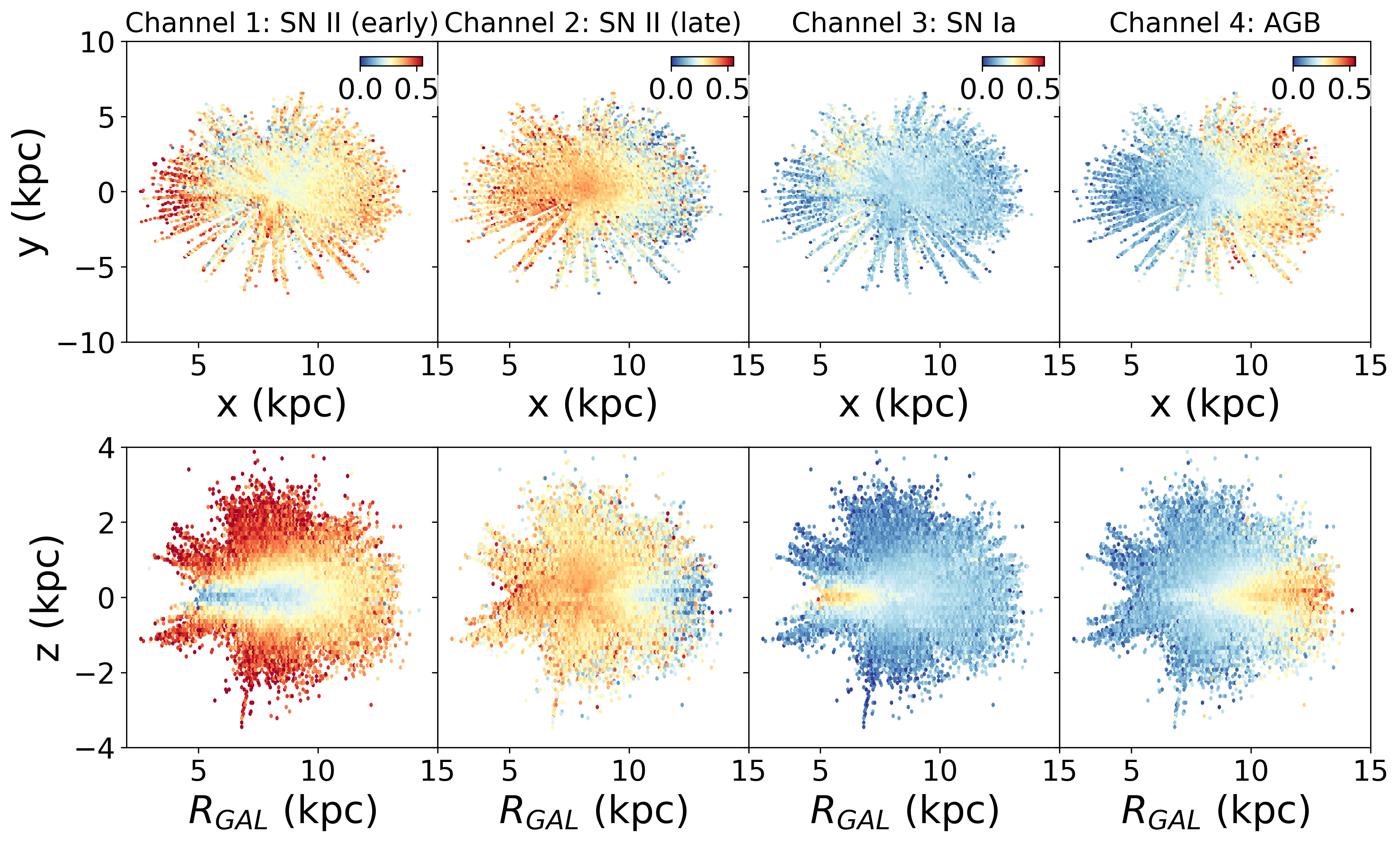}
    \caption{Spatial distribution of the fractional contribution of each latent enrichment channel across the Galactic disc.  Top row: face-on $(x,y)$ view coloured by the per-star fractional
    contribution to Channels~1--4 for stars $|z| <$ 1.5~kpc from the disc to show radial gradients in the mid-plane ($\approx 160,519 $ stars).  
    {Bottom row:} edge-on $(R,z)$ distribution for the $\approx 169,443$ stars. The colours reflect the value of the channel fraction (truncated at $f=0.5$ for clarity). Each channel exhibits clear and coherent spatial organisation: 
    Channel~1 shows its highest expression at smaller radii and larger $|z|$, 
    Channel~2 is distributed across height at inner and intermediate radii, 
    Channel~3 dominates near the mid-plane and intermediate radii, 
    and Channel~4 strengthens toward the outer disc. These patterns demonstrate that the latent abundance channels follow smooth chemical–spatial structures that reflect enrichment pathways broadly consistent with changing prevalence of enrichment by early and later core–collapse supernovae (Channels~1 and 2), delayed iron production from SNe~Ia (Channel~3),  and AGB contributions that increase with radius (Channel~4).}
    \label{fig:spatial_channels}
\end{figure*}

Similarly to Paper I, we emphasize that the nucleosynthetic interpretation is not required for the conclusions of this work. While we connect these patterns to enrichment channels based on dominant source signatures, the latent patterns are mathematical constructs that capture the dominant axes of variation in abundance patterns. They may reflect mixtures of physical sources, systematic features, or variance-driven modes that do not correspond one-to-one with distinct nucleosynthetic processes.  

Again, to compare the relative contributions of the latent enrichment patterns across the Galaxy, we convert the non-negative pattern coefficients (or weights)
$f_{im}$ into per-star normalised fractions. For each star $i$, the vector of pattern weights is normalised by its sum across all $M$ latent patterns:
\begin{equation}
\mathbf{f}_{im}
=
\frac{f_{im}}{\sum_{m'=1}^{M} f_{im'}} ,
\label{eqn:seven}
\end{equation}

where $m$ indexes the enrichment pattern and the summation runs over all patterns
for a fixed star $i$. By construction,
\begin{equation}
\sum_{m=1}^{M} \mathbf{f}_{im} = 1 .
\end{equation}

These normalised fractions $\mathbf{f}_{im}$ then represent the relative contribution
of each latent pattern to an individual star and provide a convenient
parametrisation for examining how enrichment patterns vary with Galactic
position, age, and orbital properties. All fractional pattern maps and trends
presented in subsequent figures use these per-star normalised fractions.

The subsequent analysis of how the latent fractions vary across Galactic radius, height, and orbital properties provides an external validation of their physical interpretability: any spatial or dynamical coherence reflects divisions in the expression of abundances, irrespective of whether the patterns correspond to pure enrichment sources.

\begin{figure}
    \centering
    \includegraphics[width=1\linewidth]{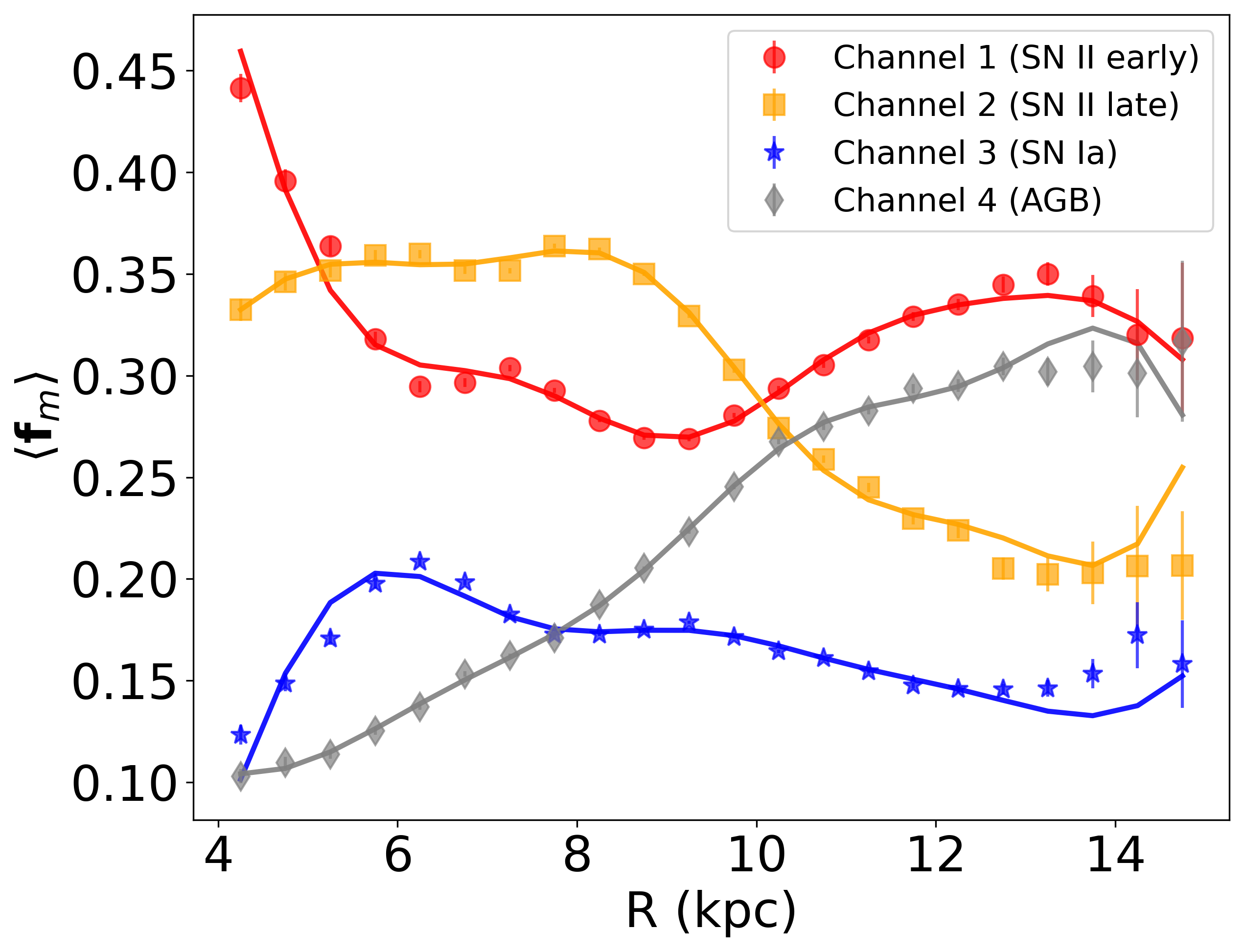}
\caption{
Mean fractional contribution of each NMF enrichment channel as a function of  Galactocentric radius for stars with $|z|< 3\,\mathrm{kpc}$. Coloured points show the  data in $0.5\,\mathrm{kpc}$ radial bins with standard-error uncertainties. The  coloured lines show the best-fitting five-mode damped radial model, in which all channels  share the same set of radial basis functions and differ only in their linear  coefficients. The model captures both the global inside--out gradient and the 
intermediate-scale curvature exhibited by the data, demonstrating that the  dominant radial structure of Galactic chemical enrichment is intrinsically  structurally coherent. The fit is not statistically perfect (it has a reduced $\chi^2 > 1$).  Nonetheless, the recovered  five-mode basis compactly describes the coherent large-scale behaviour shared by 
all enrichment channels.}
\label{fig:damped}
\end{figure}

\section{Results}

\subsection{Latent Pattern Fraction Distribution}
\label{results1}

To ensure that all chemo–spatial analyses are based on stars with reliable abundances, distances, ages, and orbital properties, we applied a sequence of quality cuts.  We restricted the sample to stars with predominantly disc phase–space coordinates by requiring Galactocentric cylindrical radius $R_{\rm gal} < 20$ kpc, vertical height from the mid-plane $|z| < 5$ kpc, and real-valued angular momentum $|L_z| < 10{,}000\ {\rm km\,s^{-1}\,kpc}$. We retained stars with positive finite age labels. We required a parallax precision better than 20\% ($\sigma_\varpi / \varpi < 0.2$) and selected only stars with abundances well generated by the latent basis using a criterion of a per-star reduced $\chi^2 < 5$ from the NMF decomposition. 

The combined selection, which enforces astrometric quality, robust age information, physically plausible spatial and orbital parameters, and well-fitted abundance patterns, yields a high-fidelity sample of approximately $169,443$ stars, which forms the set for all remaining analyses and figures in this paper.

The spatial behaviour of the four latent channels for these stars is shown in Figure ~\ref{fig:spatial_channels}.   Although the decomposition is purely data-driven, the channel fractions vary smoothly across the disc in a manner accordant with their broad nucleosynthetic associations, and as per paper~I.  Channel~1 is prominent above the mid-plane and in the inner Galaxy, consistent with the signature of enrichment dominated by early, prompt core–collapse supernovae (SN~II).  Channel~2 increases more gradually with radius and height, and  Channel~3 shows its highest fractional contribution near the mid-plane from the solar radius toward the inner Galaxy. This is consistent with expectations for SN~II and SN\,Ia-driven enrichment, respectively. Channel~4 becomes increasingly dominant toward the outer disc and at low and moderate heights, in line with stronger relative contributions from AGB-produced $s$-process material in lower-metallicity, radially extended populations. The smooth spatial organisation of all four patterns shows that the latent
channels capture coherent modes of chemical evolution rather than arbitrary mathematical components.

\subsubsection{Latent Pattern Radial Mode Decomposition}
\label{sec:radialmodes}

To quantify the radial organisation of the four NMF enrichment channels shown spatially in Figure \ref{fig:spatial_channels}, we measure how each channel’s fractional contribution varies with Galactocentric radius $R_{GAL}$. 

We select stars satisfying our global ``good'' quality criteria 
and restrict the sample to $|z|<3\,\mathrm{kpc}$ to minimise contamination  from vertically heated populations. The stars are binned in radius using  uniform $0.5\,\mathrm{kpc}$ bins. 

For each channel $m$ and radial bin $r$ we compute the mean fractional
contribution,

\begin{equation}
\mathbf{f}_m(R_r) = \frac{1}{N_r} \sum_{i \in \mathcal{R}_r} \mathbf{f}_{im},
\end{equation}

where $\mathbf{f}_{im}$ is the normalised NMF-derived latent fraction for star $i$ in channel $m$, as per Equation \ref{eqn:seven},
$\mathcal{R}_r$ denotes the set of stars whose Galactocentric radii fall within radial bin $r$ (centred at $R_r$), and $N_r$ is the number of stars in that bin. Uncertainties are estimated as the standard error of the mean,
$\sigma_{m,r}=\mathrm{std}(\mathbf{f}_{im})/\sqrt{N_r}$. We enforce a minimum of 20 stars per bin. These bins of fractional contribution for each channel are shown in Figure \ref{fig:damped}. 

We observe clear structure in the fractional contributions across radius, arising partly from the Milky Way Mapper selection function but also plausibly reflecting astrophysical structure such as radial migration, perturbations, and population–mixing effects. This structure changes under different selection of $|z|$ and age; therefore we focus not on the specific characteristics but are interested the joint behaviour.  The four fractional contributions for each star must sum to unity; however, this constraint alone does not impose spatial coherence among the channels across radius. To assess whether their behaviour is governed by a shared underlying driver, we test whether all four channels can be described by a single, flexible low-dimensional radial basis. A successful fit is not guaranteed: it would indicate that the enrichment channels respond coherently to the same global evolutionary modes, rather than exhibiting unrelated spatial structure.

We model the radial behaviour of the enrichment channels using a flexible basis,
by expressing each channel as a linear combination of a number of global damped
radial modes. Motivated by the smoothness of the observed trends and the
expectation that large-scale chemical evolution is governed by only a few global
degrees of freedom, we adopt a basis of the form
\begin{equation}
\Phi_d(R) = \exp(-a_d R)\,\cos(b_d R),
\end{equation}
with $D=5$ modes, where $d=1,\dots,D$, which we find sufficient to capture the
observed radial structure. The parameters $\{a_d,b_d\}$ define a shared basis common to all
channels. Each NMF channel $m$ is then represented as
\begin{equation}
f_m(R) \;\approx\; \sum_{d=1}^{D} A_{md}\,\Phi_d(R),
\end{equation}
where the amplitudes $A_{md}$ are channel-specific linear coefficients.

The shared basis parameters $\{a_d,b_d\}$ are determined by minimising the total
weighted residual across all channels,
\begin{equation}
\chi^2(\{a_d,b_d\}) =
\sum_{m}\sum_{r}
\frac{
\left[
f_m(R_r) - \sum_{d=1}^{D} A_{md}\,\Phi_d(R_r; a_d,b_d)
\right]^2
}
{\sigma_{m,r}^2},
\end{equation}
where, for any fixed choice of $\{a_d,b_d\}$, the channel-specific coefficients
$A_{md}$ are obtained analytically via weighted linear least squares. This
procedure yields a single shared set of radial modes that span the large-scale
radial variation of all enrichment channels.

The damped-mode basis is best constrained in the mid-disc (6--12 kpc). Although increasing the number of modes $D$ monotonically improves the formal fit, fitting the largest radii and reducing both the Akaike Information Criterion (AIC) and the Bayesian Information Criterion (BIC)
 up to $D \approx 9$ ,  five modes capture the large-scale and intermediate-scale radial trends.  We therefore adopt $D=5$ as the meaningful indicator of overall behaviour of the data.

The extracted basis demonstrates that the radial behaviour of all four enrichment channels can be described using the same small set of global radial modes, with each channel differing only in the relative weighting of those modes. This demonstrates that their large-scale evolution is constrained and coherently coupled, rather than independently smooth.
In the Appendix (Figure~\ref{fig:residuals}, we show residual maps in the Galactic $x-y$ plane after subtracting the shared radial mode model. The remaining structure suggests azimuthal inhomogeneity in the relative contributions of the enrichment channels. These residuals are comparable in amplitude to the residuals seen in [Fe/H] and other abundances after subtraction of a smooth radial model in \citet{Hawkshaw2024}. The enrichment-channel residuals suggest that the disc is not fully axisymmetric in its mixture of enrichment pathways at fixed radius. However, additional effects, including the selection function, stellar age, and height above the plane, should be modelled before interpreting this structure in detail. Larger Milky Way Mapper samples with corresponding ages will enable this signal to be investigated further.

\begin{figure*}
    \centering
    \includegraphics[width=1\linewidth]{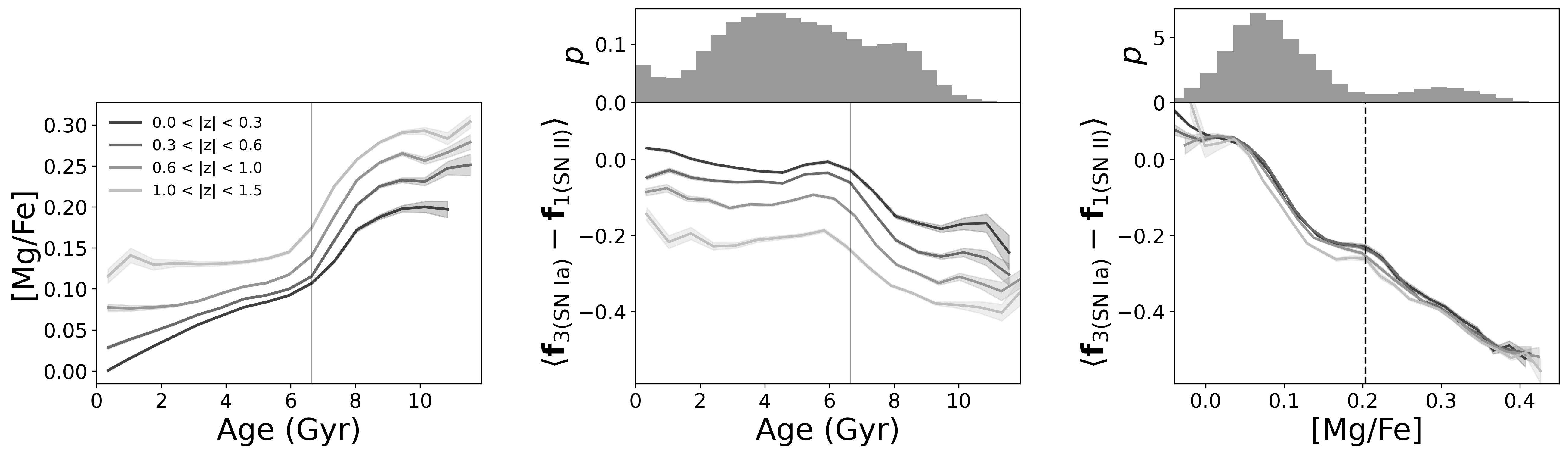}
    \caption{The evolution of the abundance plane and latent enrichment fractions as a function of age and height above the Galactic plane. The bins in $|z|$ (in kpc) corresponding to the different lines are indicated in the left-most panel. Left: Running mean of [Mg/Fe] versus stellar age, shown separately in bins of vertical height $|z|$. Middle: Difference between the latent fractions associated with the SN~Ia- and SN~II-dominated channels as a function of age; the histogram above shows the age distribution of the sample. Right: The same fractional difference shown as a function of [Mg/Fe], with the histogram above showing the [Mg/Fe] distribution. In both the left and middle panels, the age corresponding to the maximum rate of change in the trend at smallest $|z|$ bin  is indicated, occurring at $\sim$6~Gyr. In the right panel, the location of the valley in the $\alpha$-bimodality is marked, where the fractional difference deviates from a smooth trend. Together, these panels demonstrate that both the abundance plane and the latent enrichment fractions independently identify a common transition in disc chemical evolution.}
    \label{fig:transition}
\end{figure*}

\subsubsection{Transitions in Enrichment}

We examine whether the latent enrichment fractions can be used to identify transitions in disc chemical evolution, and whether they provide additional insight into the relationship between the low- and high-$\alpha$ populations. We begin with the familiar abundance measurements. In the left panel of Figure~\ref{fig:transition}, we show the running mean of [Mg/Fe] as a function of stellar age, computed separately in bins of vertical height bins above the plane, $|z|$. The point of maximum rate of change in these trends is indicated. Across all height bins, the [Mg/Fe]-age relation steepens  at ages older than $\sim$6\,Gyr and then flattens again toward $\sim$10\,Gyr, signaling a change in the pace of chemical evolution at these epochs.

A consistent transition is independently recovered in the latent enrichment space. The middle panel of Figure~\ref{fig:transition} shows the difference between the latent fractions of channels~3 and~1, corresponding approximately to delayed (SN~Ia-dominated) and prompt (SN~II-dominated) enrichment, respectively. Negative values of this fractional difference indicate a larger contribution from the SN~II-like channel, while positive values indicate a reduced fractional contrast between the two channels. The age at which the rate of change in this fractional difference is maximal coincides with the same $\sim$6\,Gyr epoch identified in the [Mg/Fe]--age relation, indicating a transition in the relative balance of enrichment sources.

In the rightmost panel, the same fractional difference is shown as a function of [Mg/Fe]. Rather than following a smooth monotonic trend, the relation deviates near the valley of the $\alpha$-bimodality, where the fractional contrast between enrichment channels changes most rapidly. The location of the valley is marked, and the age and [Mg/Fe] distributions of the sample are shown above the middle and right-hand panels for reference.

These figures demonstrate that the latent enrichment fractions capture coherent transitions in disc chemical evolution that are visible across multiple projections of the data. The agreement between the abundance-plane behaviour and the latent-channel diagnostics indicates that these transitions reflect genuine changes in enrichment conditions. The latent representation therefore provides a complementary and information-rich way to identify and characterise the  key phases in the Milky Way’s enrichment history.

\subsection{Grouping Stars by Pattern Fraction}
\label{results2}

To identify coherent groups of stars with similar mixtures of enrichment patterns, we apply \textit{k}-means clustering to the NMF-derived pattern fractions. Because the per-star offset removes the absolute abundance scale, the clustering operates on the \emph{shape} of each star’s abundance pattern. That is, its relative element-to-element structure, rather than on metallicity. This choice is important; using the per-star offset suppresses this metallicity-driven structure, and reveals the second-order pattern differences that are physically diagnostic of similar underlying enrichment conditions. This allows the clustering to identify stars with different absolute abundances but genuinely similar enrichment \emph{patterns}, which is the quantity of astrophysical interest here.

In this context, $k$ denotes the number of clusters specified a priori; setting $k=1500$ partitions our $\approx 169,443$ stars into an average of 112 stars per group (and a standard deviation of 50 stars), giving clusters of reasonably comparable size and statistical fidelity, while still preserving diversity in enrichment between the groups. Although the precise choice of $k$ is arbitrary, this value provides a balance between granularity and statistical robustness. We verify that the same overall trends are recovered for other values of $k$ (e.g., $k=100, 1000$), although at lower granularity. We also highlight that these groups do not, by design, contain chemically identical stars, but stars with similar abundance patterns. In the following sections, we analyse the properties of these clusters in age, chemistry, and kinematics, to understand the physical nature of the enrichment paths they represent in terms of the chemical evolution of the Milky Way disc.

\begin{figure*}
    \centering
\includegraphics[width=1\linewidth]{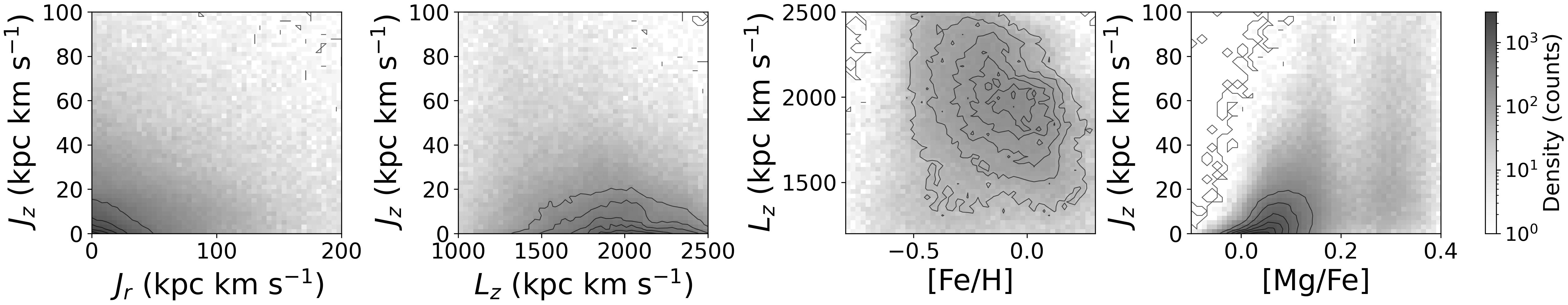}
\includegraphics[width=1\linewidth]{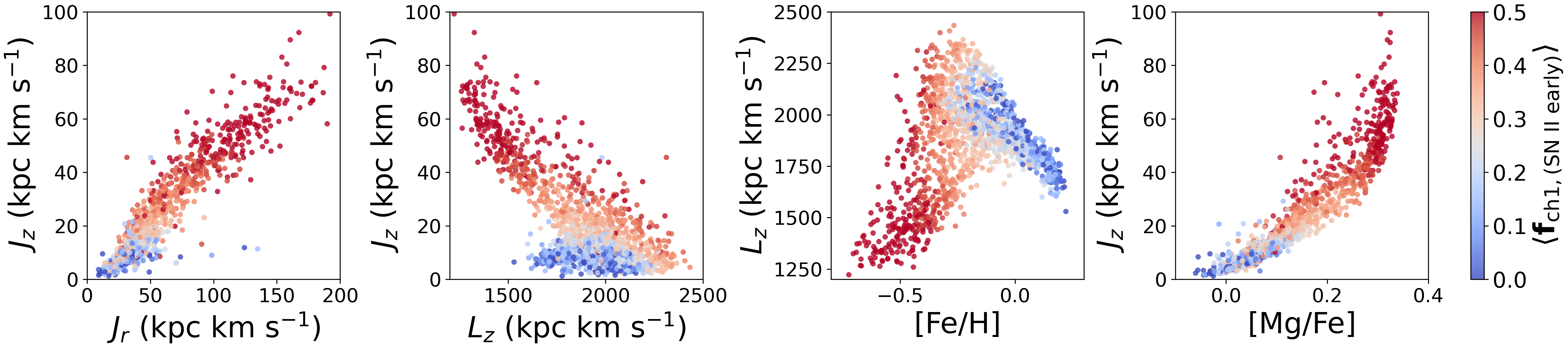}
\includegraphics[width=1\linewidth]{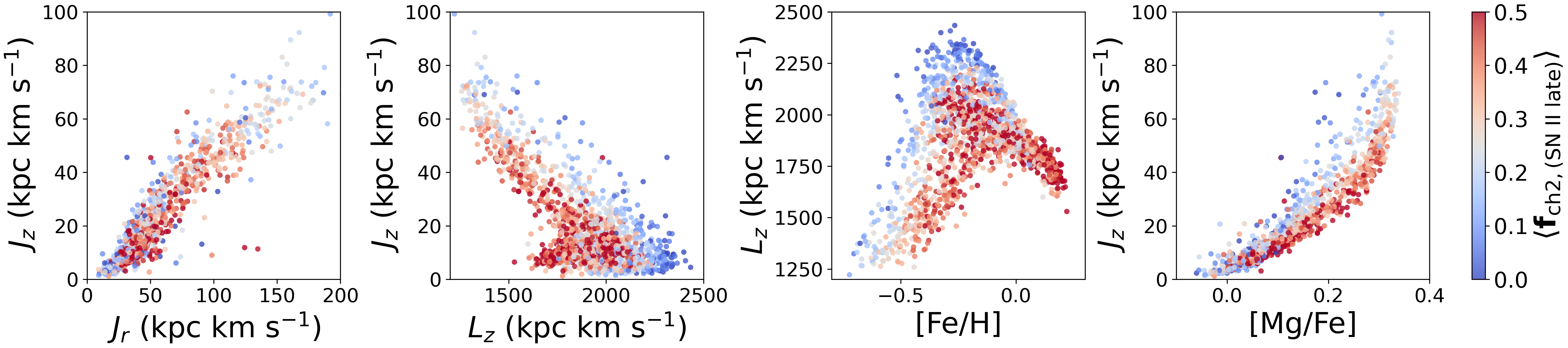}
\includegraphics[width=1\linewidth]{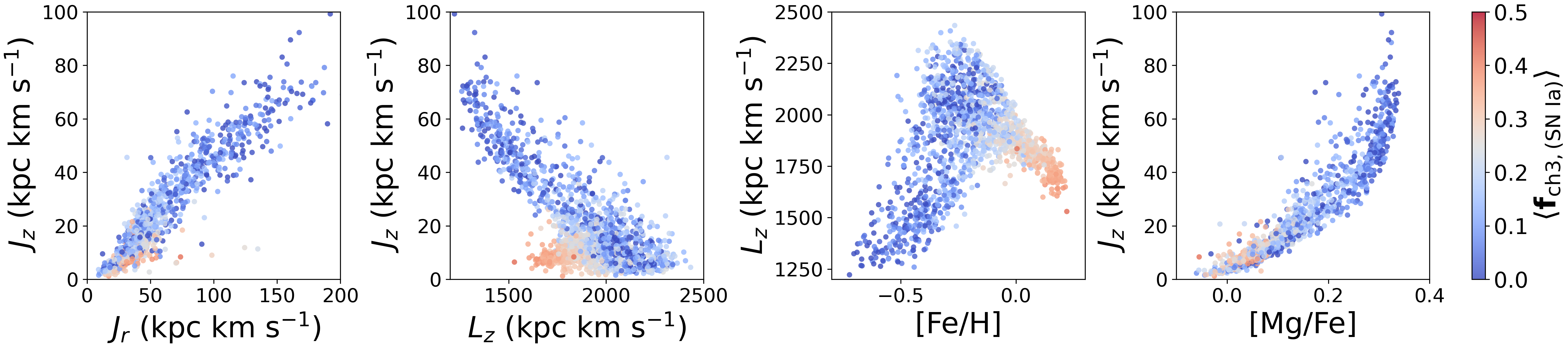}
\includegraphics[width=1\linewidth]{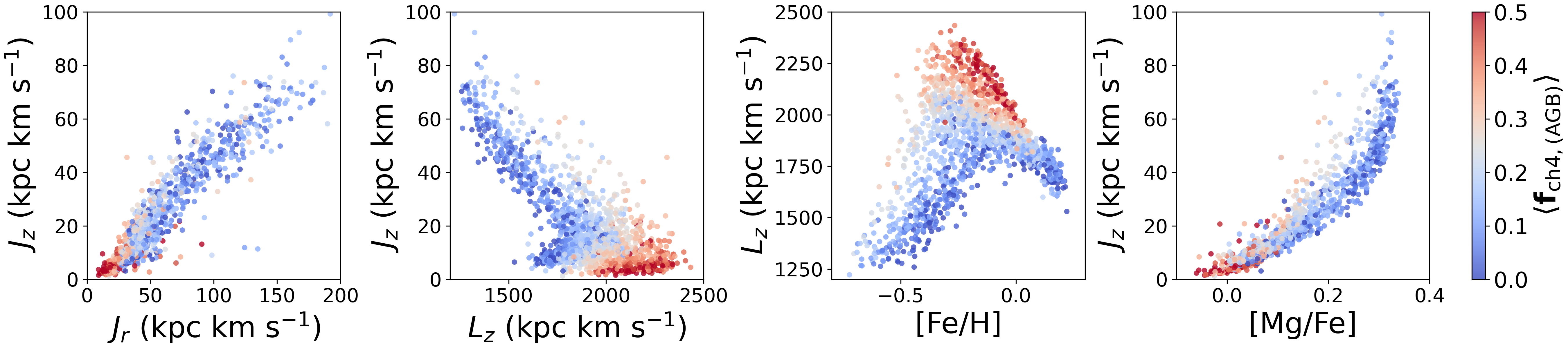}
    \caption{Chemo-dynamical properties of the 1,497 stellar groups defined by their NMF latent pattern fractions (rows 2–5), contrasted with the background density distribution of all stars used to construct the groups (top row). Each group is coloured by its fractional contribution to one representative latent channel: $f_{\mathrm{ch}1}$ (second row), $f_{\mathrm{ch}2}$ (third row), $f_{\mathrm{ch}3}$ (fourth row), and $f_{\mathrm{Ch}4}$ (fifth row). These channels correspond broadly to enrichment from early and late prompt sources (SN~II) and delayed sources (SN~Ia, AGB), respectively. The groups trace a set of discrete, ordered evolutionary pathways in both enrichment and dynamical properties, reflecting the coherent chemo-dynamical structure captured by the latent-space decomposition.}
    \label{fig:latent_groups}
\end{figure*}

\begin{figure*}
    \centering
    \includegraphics[width=1\linewidth]{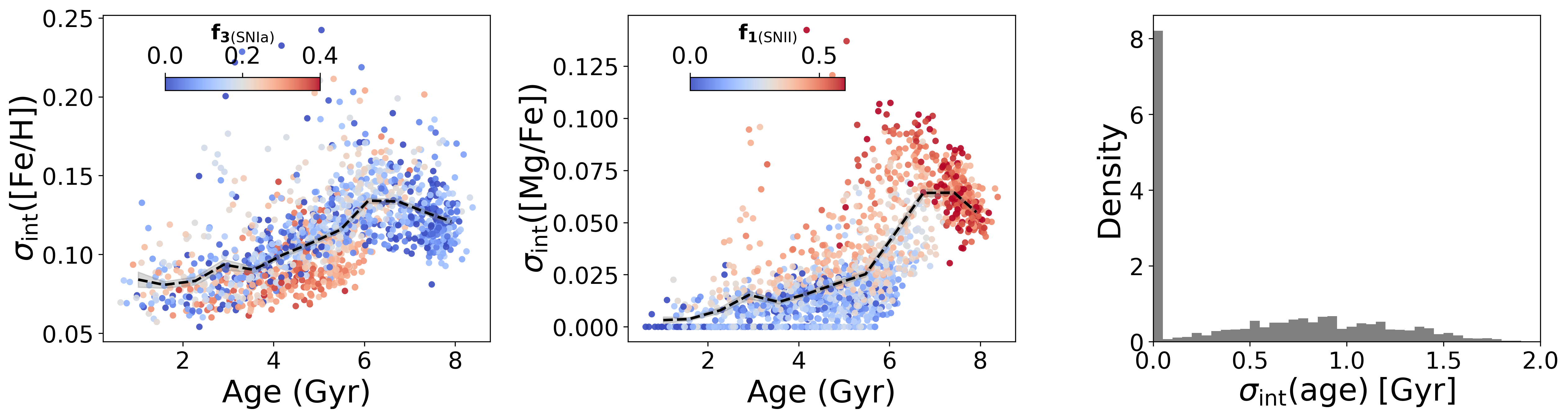}
    \caption{The 1,497 k-means clusters are shown as a function of their mean stellar age. The left and middle panels display the intrinsic abundance dispersion within each cluster in [Fe/H] and [Mg/Fe], respectively, after accounting for measurement uncertainties. Clusters are colour-coded by the mean contribution of the associated enrichment channel. In both abundance dimensions, the intrinsic dispersion increases at ages of $\sim$6 Gyr, reaches a maximum near $\sim$6–7 Gyr, and subsequently declines at older ages. The right-most panel shows the histogram of intrinsic age dispersion; the majority are consistent with zero intrinsic age spread, indicating that clusters are typically narrow in age.}
   \label{fig:feh_dispersion_channels}
\end{figure*}

\subsubsection{Clustered Groups of Similar Latent Patterns}

We first examine whether the $k$-means groups defined in latent-fraction space
trace coherent dynamical and chemical structures in the Milky Way disc.
Figure~\ref{fig:latent_groups} shows the median values of the groups across several
dynamical–abundance projections, coloured by their median fractional
contribution of enrichment channels~1-4, corresponding to early and late SN~II, 
SN~Ia and AGB enrichment, respectively. Groups with fewer than ten stars are excluded, leaving 1497 groups in total.

From left to right, the panels show the radial–vertical action plane
($J_r$ versus $J_z$), the vertical action–angular momentum plane
($J_z$ versus $L_z$), the angular momentum–metallicity plane
($L_z$ versus [Fe/H]), and the vertical action–$\alpha$-abundance plane
($J_z$ versus [Mg/Fe]). The top row of Figure~\ref{fig:latent_groups} displays
the stellar density distribution in each projection, providing the global
context. The rows below show the same projections coloured by the median
fractional contribution of each enrichment channel.

The stark contrast between the density maps and the channel-coloured panels
demonstrates that the latent groups do not simply recover regions of high
stellar density or isolated islands in phase space. Instead, they trace
continuous sequences across all projections. Although only the group medians
are shown, each group spans a wide range in phase-space and abundance
coordinates, as indicated by the underlying density distribution in the top
row.

In the $J_r$–$J_z$ plane, the groups form a tight, ordered sequence: clusters
with larger vertical excursions systematically exhibit higher fractions of
channel~1 (SN~II, early) enrichment, while groups confined closer to the disc plane are
more strongly enriched by channels~3 and  4 (SN~Ia and AGB). The correlation between $J_z$ and enrichment channel~1 (SN~II, early) indicates that the vertical structure of the disc retains a strong imprint of its chemical history that is captured in the groups. 

The $L_z$–$J_z$ plane reveals two distinct but connected regimes. Dynamically cold groups at low $J_z$ transition smoothly from SN~Ia–dominated enrichment at smaller $L_z$ to increasing SN~II contributions at larger $L_z$. In contrast, dynamically hotter groups follow a correlated decrease in $L_z$ with increasing $J_z$ and are
strongly dominated by SN~II enrichment. The detailed structure of the [Fe/H]-$L_z$ plane is influenced by the radial selection function of the sample. Nevertheless, the latent groups retain a clear ordered relationship between abundance pattern, metallicity, and angular momentum. The structure in the [Fe/H]–$L_z$ plane can be interpreted as a projection of the
classical thin–thick disc dichotomy. At fixed angular momentum, stars populate
two metallicity sequences, corresponding to a metal-poor and a metal-rich
branch. Within the latent-fraction framework, the metal-poor sequence is
dominated by enrichment from channel~1 (SN~II, early), while the metal-rich sequence is
dominated by channel~3 (SN~Ia). These two branches converge at intermediate metallicity, where we observe a transition in channels~1 (SN~II, early) and ~3 (SN~Ia). Notably, channels~2 (SN~II, late) and ~4 (AGB) do not show a similar transition at the location where these branches meet. These have consistent amplitudes across the branches, with contribution fraction divided instead by the angular momentum coordinate.  Along the high-$\alpha$ branch, the mean [Fe/H] of the groups decreases toward lower $L_z$. This reflects the older, dynamically hotter structure of the high-$\alpha$ disc (see the [Mg/Fe]-$J_z$ panel at right). The low-$\alpha$ branch shows a different behaviour, with the most metal-rich groups bending back toward lower $L_z$, and the metal-rich stars associated with the inner disc. Since each point represents the median properties of stars grouped by similar latent abundance patterns, these trends show that the latent enrichment groups preserve information about the coupled chemical and dynamical structure of the Galactic disc that is set at birth. This behaviour provides a new view of the thin–thick disc bifurcation, revealing underlying enrichment changes and a connected although changing trajectory of enrichment groups, without requiring discrete or chemically independent populations.

The fourth column, showing [Mg/Fe] versus $J_z$, demonstrates that a group’s vertical dynamical structure is closely coupled to its $\alpha$-abundance. Groups with larger vertical action exhibit enhanced [$\alpha$/Fe] and higher channel~1 fractions (SN~II, early), whereas groups confined closer to the disc plane show lower [$\alpha$/Fe] and stronger contributions from channel~3 (SN~Ia). Looking at this plane for channel 2 (SN~II, late) shows that at fixed [Mg/Fe], the range in $J_z$ corresponds to a strong gradient in the channel fraction. A more mild gradient is similarly seen across intermediate [Mg/Fe] ranges for channel 1 (SN~II, early). These trends again highlight the tight coupling between vertical disc structure and chemical enrichment history.

Overall, Figure \ref{fig:latent_groups} shows that latent-fraction groups serve as building block tracers of the disc, connecting enrichment to structure. While some correlations seen here, for example, between $\alpha$-enhancement and disc kinematics, are well established, Figure~\ref{fig:latent_groups} extends this picture by showing that \emph{all} enrichment channels—including those associated with delayed sources—exhibit coherent and structured behaviour in action–abundance space. In particular, channels~2 (SN~II, late) and~4 (AGB) display systematic trends that are not trivially inferred from classical $\alpha$-based diagnostics.

\subsection{Homogeneity of Clusters}

We now examine a diagnostic of the homogeneity of the star-forming environment over time by evaluating the intrinsic dispersion of [Fe/H] and [Mg/Fe] within groups as a function of their mean age. Because the latent channels are defined
in relative abundance space, with the absolute metallicity scale removed (Equation~5), stars sharing the same channel composition are not required to have the same absolute abundances. As a result, the dispersion in [Fe/H] and [Mg/Fe]
within each group provides a measure of the range of enrichment levels sampled by a given chemical pattern at a given epoch.

This is shown in Figure \ref{fig:feh_dispersion_channels} (left and middle panels). In the right panel the histogram of the intrinsic age dispersion is shown. The intrinsic dispersion in [Fe/H] and [Mg/Fe] and age is the scatter accounting for the measurement uncertainty, estimated for each group as; 

\begin{equation}
\begin{aligned}
\sigma_{\mathrm{int}}([\mathrm{Fe/H}]) &= 
\sqrt{ \sigma_{[\mathrm{Fe/H}]}^{2} - \left\langle \mathrm{err}_{[\mathrm{Fe/H}]} \right\rangle^{2} }, \\[6pt]
\sigma_{\mathrm{int}}([\mathrm{Mg/Fe}]) &= 
\sqrt{ \sigma_{[\mathrm{Mg/Fe}]}^{2} - \left\langle \mathrm{err}_{[\mathrm{Mg/Fe}]} \right\rangle^{2} }, \\[6pt]
\sigma_{\mathrm{int}}(\mathrm{Age}) &= 
\sqrt{ \sigma_{\mathrm{Age}}^{2} - \left\langle \mathrm{err}_{\mathrm{Age}} \right\rangle^{2} } .
\end{aligned}
\end{equation}

Notably, no latent-fraction group is dominated by the oldest stars. Although stars with ages exceeding $8.7$~Gyr are present in the sample, they are distributed across multiple latent groups rather than forming a single, chemically distinct cluster. This indicates that the earliest phases of disc enrichment represented by the inferred latent abundance patterns do not correspond to a chemically isolated population. Instead, stars at the oldest ages share enrichment patterns with stars at higher metallicity. 

In Figure~\ref{fig:feh_dispersion_channels}, the intrinsic dispersion in both [Fe/H] and [Mg/Fe] reaches a clear maximum at $\sim$6--7~Gyr. This peak identifies an epoch during which stars sharing similar latent enrichment patterns span the
largest range in absolute [Fe/H] and [Mg/Fe] values. This epoch coincides with a change in the relative contribution of channel~1 (SN~II, early).  Because the latent channels are defined in relative abundance space, with the absolute metallicity scale removed, a large spread in [Fe/H] or [Mg/Fe] at fixed channel composition reflects variation in the overall level of enrichment rather than differences in the underlying nucleosynthetic pattern. The increased dispersion therefore indicates that a common set of enrichment patterns is realised across a broader
range of enrichment levels at these ages. 
The right-hand panel of Figure~\ref{fig:feh_dispersion_channels} further shows that many latent-fraction groups are consistent with being mono-age populations, with $48\%$ exhibiting no measurable intrinsic age dispersion and the remainder having a median intrinsic dispersion of $\langle\sigma_{\mathrm{age,int}}\rangle = 0.8$~Gyr.

 \begin{figure*}
\includegraphics[width=1\linewidth]{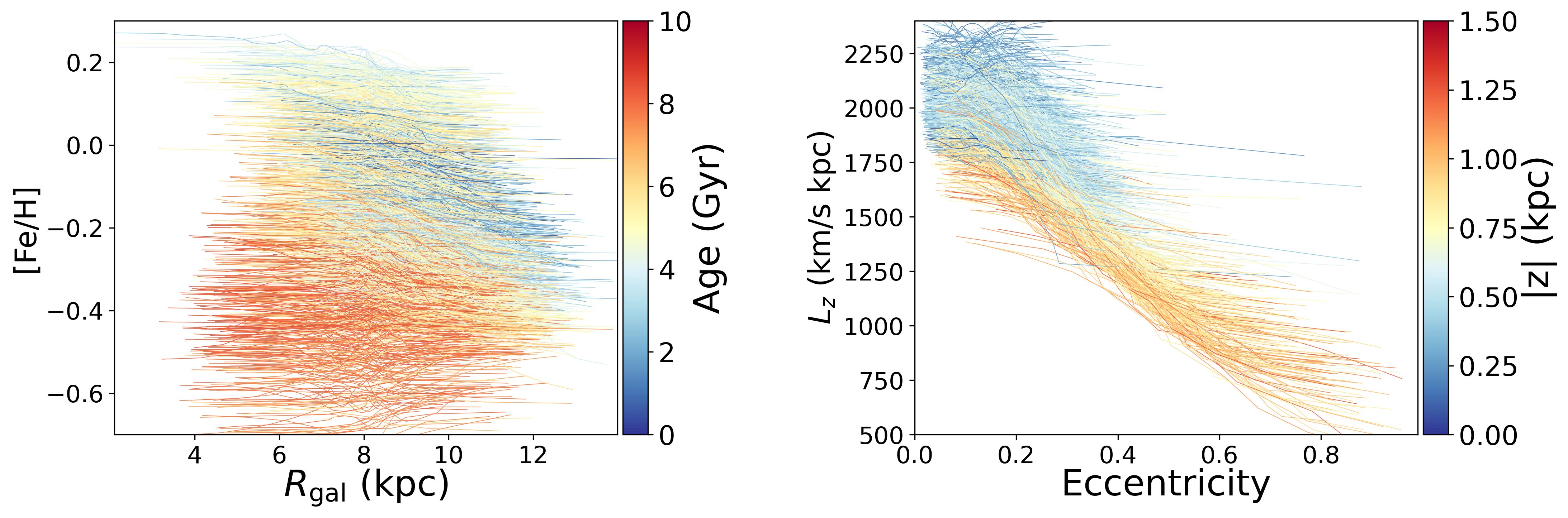}
    \caption{The spatial and dynamical behaviour of the empirical enrichment paths of the k-means clusters. Left: paths of [Fe/H] versus Galactocentric radius $R_{\mathrm{gal}}$, colored by age. There is a strong correlation between the trajectory gradients and age and an inversion of the gradient for older stars. Right: Angular momentum $L_z$  versus eccentricity, colored by mean $|z|$,  revealing dynamical structure, whereby eccentricity evolves from small to large values for paths with stars at larger mean distances from the mid-plane in the inner Galaxy at smaller $L_z$. This figure illustrate how these k-mean cluster pathways of stars with similar abundance patterns, are closely tied to spatial and orbital distributions. }
    \label{fig:pathsspatial2}
\end{figure*}

\subsection{Empirical paths across spatial dimensions}
\label{ce}

Our empirically defined groups are collections of stars that share similar latent abundance patterns. We have examined how the groups show evolving and correlated mean abundance, spatial and orbital action distributions by looking at their mean properties and the correlations between these properties.  We now examine the spatial and dynamical structure of the stars in each group. 

Figure~\ref{fig:pathsspatial2} shows representative projections of stars within each latent-fraction group across the $R_{\mathrm{gal}}$–[Fe/H] plane and the eccentricity–$L_z$ plane. For each group, a single track is constructed as a running mean trend of the $y$-axis variable as a function of the $x$-axis variable. Stars are first ordered by the $x$ coordinate and binned into intervals containing an equal number of stars (five per bin). The mean $y$ value is computed within each bin, and the resulting sequence of bin means is smoothed using a one-dimensional Gaussian filter with width $\sigma = 2.5$ bins, applied along the bin index. This smoothing suppresses small-scale noise while preserving the large-scale structure of the trends, and corresponds to averaging over approximately 15 stars.

We emphasise that the track of each group is not tracing evolution over time. Rather, the evolution across all groups together traces evolutionary paths and are subsequently coloured by their median age to show this. These tracks each connect stars with similar abundance patterns and therefore, correspond to similar underlying enrichment conditions across time/location. The structure in these planes captures when and where those enrichment conditions operated in the disc, plus subsequent dynamical evolution.

The left panel of Figure~\ref{fig:pathsspatial2} shows that each group traces a shallow trajectory across radius, indicating that stars of a given mean age and latent-abundance composition are drawn from a broad range of Galactocentric radii. The weak radial gradients of the paths reflect the fact that mono-age populations in the Milky Way are not confined to a single birth radius: radial mixing disperses stars across the disc, so groups with similar chemical ``fingerprints'' naturally span several kiloparsecs. 

The slope of the radial metallicity gradient varies systematically with mean age. At older ages ($\gtrsim 6$~Gyr), most groups show flat or weakly positive slopes  and younger stars show negative slopes. This  captures the formation and subsequent evolution of the disc in these paths. At young ages the tracks more closely follow the present-day interstellar medium gradient: stars in the outer disc are more metal-poor. At old ages, however, the relation flattens and in some cases reverses. This inversion is a natural outcome of radial migration acting on a population born when the ISM gradient was weak \citep{Schonrich2017}. The more metal-rich stars formed in the inner disc have been redistributed to larger radii over time, while metal-poor stars formed at larger radii have remained closer to their birth locations. 

Across all NMF clusters, the distribution of stars in 
$(L_z,$ eccentriciy) space in Figure \ref{fig:pathsspatial2} shows that the groups are organised into an evolutionary path where angular momentum declines smoothly with increasing eccentricity,  indicating sorting of stars into progressively smaller guiding radii and more radial orbits.  The colour coding by $|z|$ reveals the coupled vertical–orbital structure  of the disc, with stars at larger $|z|$ occupying the lower-$L_z$, higher-eccentricity end of the sequence.  The continuous path traced by each latent group across the dynamical plane demonstrates that the NMF-derived abundance structure does not isolate distinct dynamical populations, but rather distinct chemical ones whose birth orbits and subsequent evolution have left coherent imprints.

\begin{figure*}
    \centering
\includegraphics[width=1\linewidth]{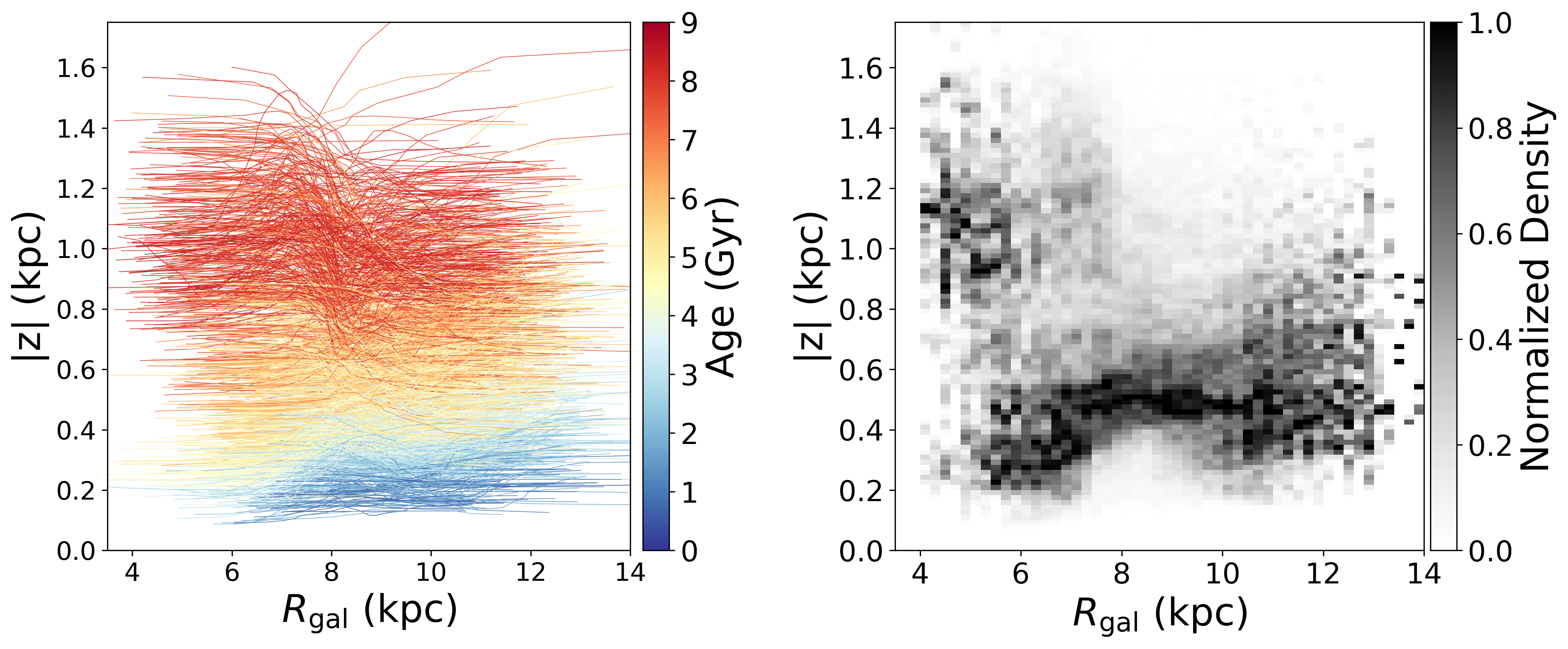}
\includegraphics[width=1\linewidth]{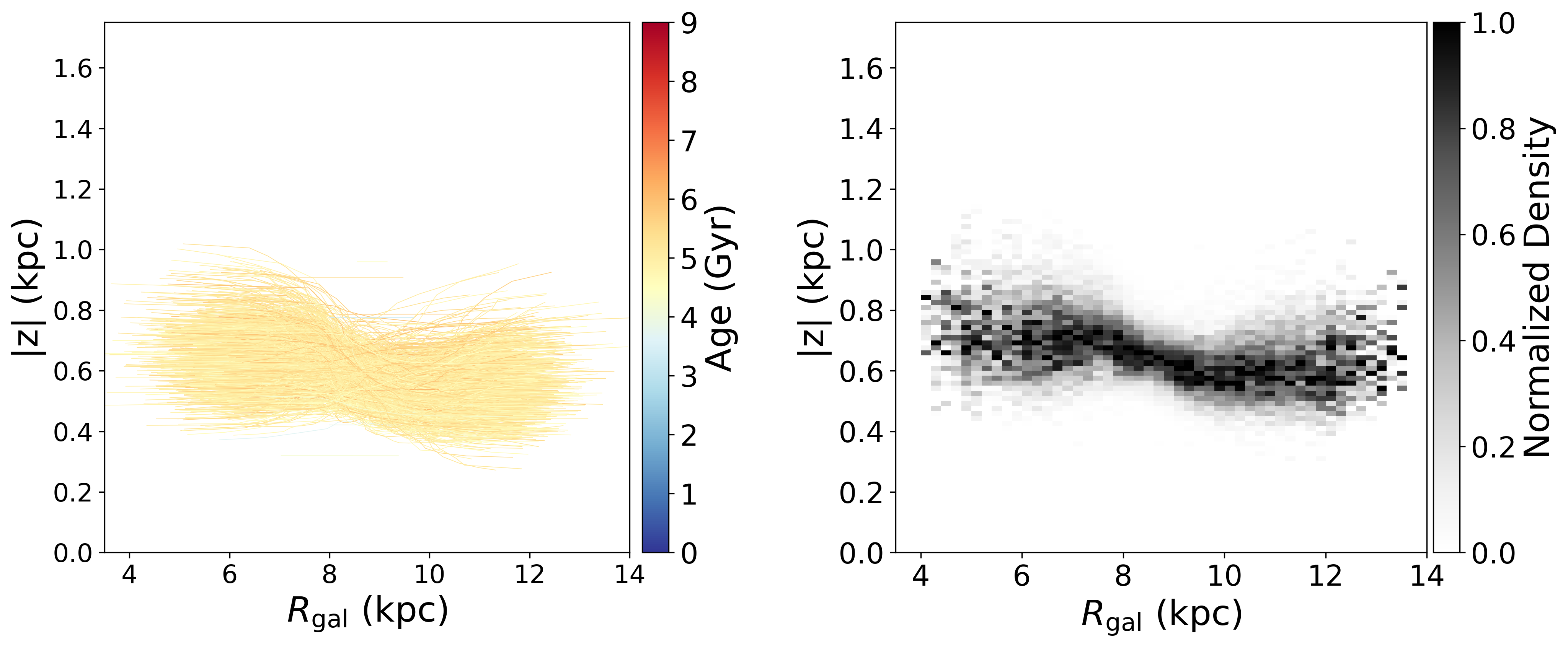}
   \caption{
    Top Left: Vertical–radial paths of the k-means groups in the Galactic disc, colored by median age. Top Right: Column density normalised map of these paths in the $R_{gal}$--$|z|$ plane to emphasise the path structure. The mean age-colored paths reveal coherent structure, including a dip-shaped feature around solar radii ($R_{\mathrm{gal}} \gtrsim 8-9$~kpc) in old stars, seen as a suppression of vertical extension at right, and vertical stratification by age.  Older groups occupy higher vertical heights, while younger groups remain confined closer to the mid-plane. There is evidence of disc flaring in the paths of younger stars closer to the plane at larger radii. Bottom panels: Same as the top panels, but with cluster labels randomly permuted, preserving cluster sizes: this provides a null test that reflects only the survey selection function and the global Milky Way disc density, independent of latent chemical similarity.}
    \label{fig:disc_flaring}
\end{figure*}

Figure~\ref{fig:disc_flaring} shows, in the top panels, the running mean of the stars in each group, across the $R$--$|z|$ plane. The panel at left traces tracks colored by mean age. The panel at right shows the column normalized stellar density distribution of the tracks. The age-colored tracks again display clear structure: older populations are found at higher mean $|z|$, and younger populations are more confined to the mid-plane, with flaring at larger radius. Together these tracks trace the evolutionary path of the Galactic disc over time. The tracks in the right-hand panel show the mean vertical undulations of the disc, with minima at approximately 6 and 10~kpc and a maximum near 8.5~kpc, while the overall thickness of the disc decreases beyond the solar radius. This undulating behaviour disappears when stars are grouped randomly, as shown in the bottom panels. The bottom panels provide a null test in which cluster membership is randomly permuted, preserving the stellar sample and survey selection function but removing any association with latent chemical similarity. The absence of undulations in this case demonstrates that the signal is not a selection effect, but instead reflects a coherent dynamical signature preserved within chemically defined populations. In the randomised case, the density distribution and tracks of $|z|$ as a function of radius reflect only the global disc structure, with the thick disc dominating the inner Galaxy (higher mean $|z|$) and the thin disc extending to larger radii: all organised temporal and finer-grained spatial structure, such as the distinct minima at 6 and 10.5~kpc, disappears.

\begin{figure}
    \centering
    \includegraphics[width=1\linewidth]{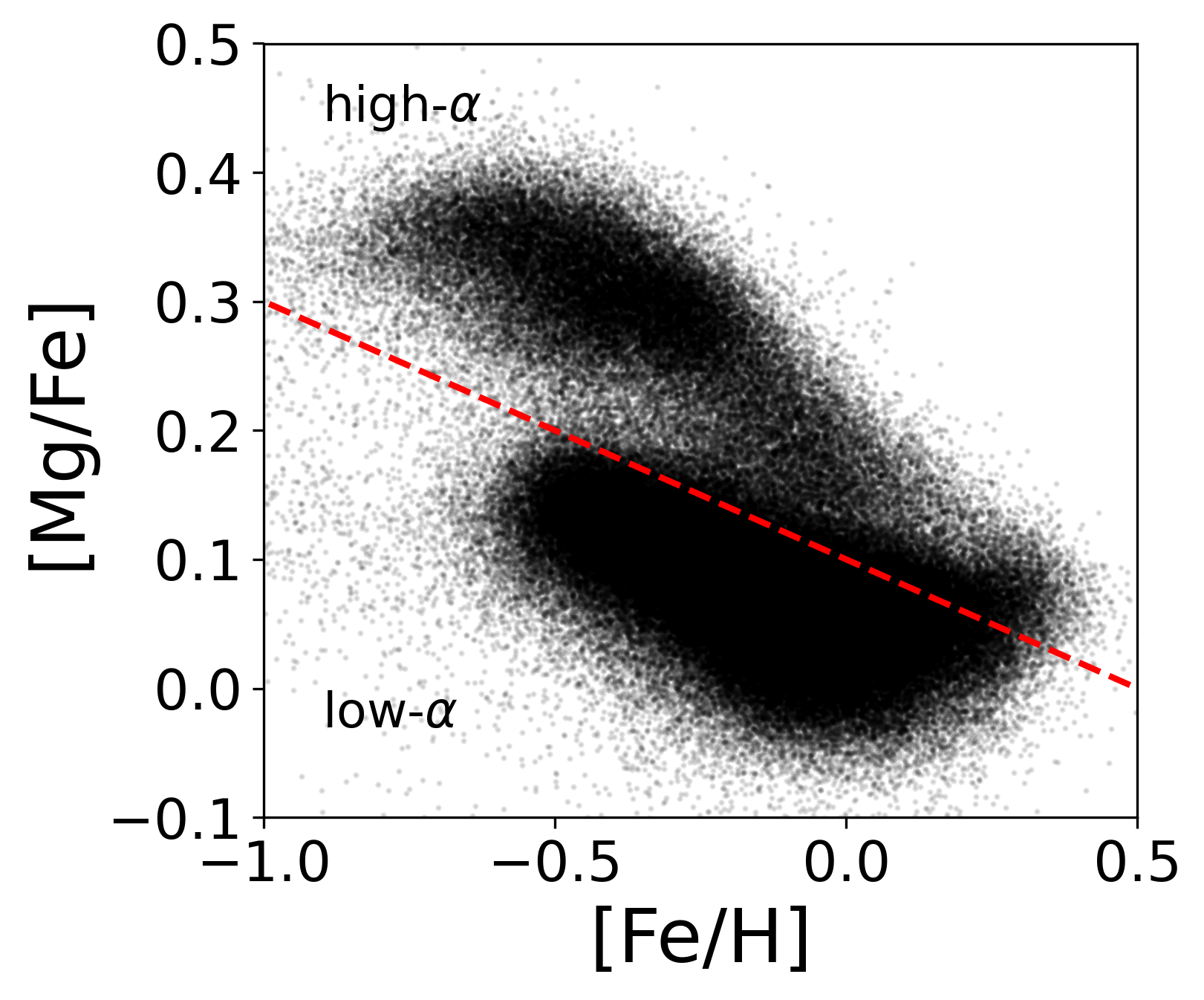}
\caption{
The $\approx 199,290$ disc stars in the [Fe/H]–[Mg/Fe] plane, illustrating the division between the low-$\alpha$ (137,624 stars) and high-$\alpha$ (61,666 stars) populations. 
The NMF basis in this subsection is learned using \textit{only} the low-$\alpha$ stars (below the dividing line). These learned patterns are then applied to the high-$\alpha$ stars to infer their latent fractions and to test how well the low-$\alpha$ chemical basis can reproduce the abundances of the high-$\alpha$ disc. 
}
 \label{fig:selection}
\end{figure}

\subsection{Latent groups separated by $\alpha$-enrichment}

To test the connection between the high- and low-$\alpha$ disc within this framework, we apply a conventional high/low-$\alpha$ division to the latent clusters. Each cluster is assigned to the high-$\alpha$ or low-$\alpha$ population based on its mean abundance values, using the dividing line shown in Figure~\ref{fig:selection}. We then examine the resulting populations in spatial coordinates and orbital action space (as per Figure~\ref{fig:latent_groups}). This approach tests whether clusters that overlap in dynamical and spatial properties also share similar latent chemical fractions, despite differences in $\alpha$-enhancement.

\begin{figure*}
    \centering
    \includegraphics[width=1\linewidth]{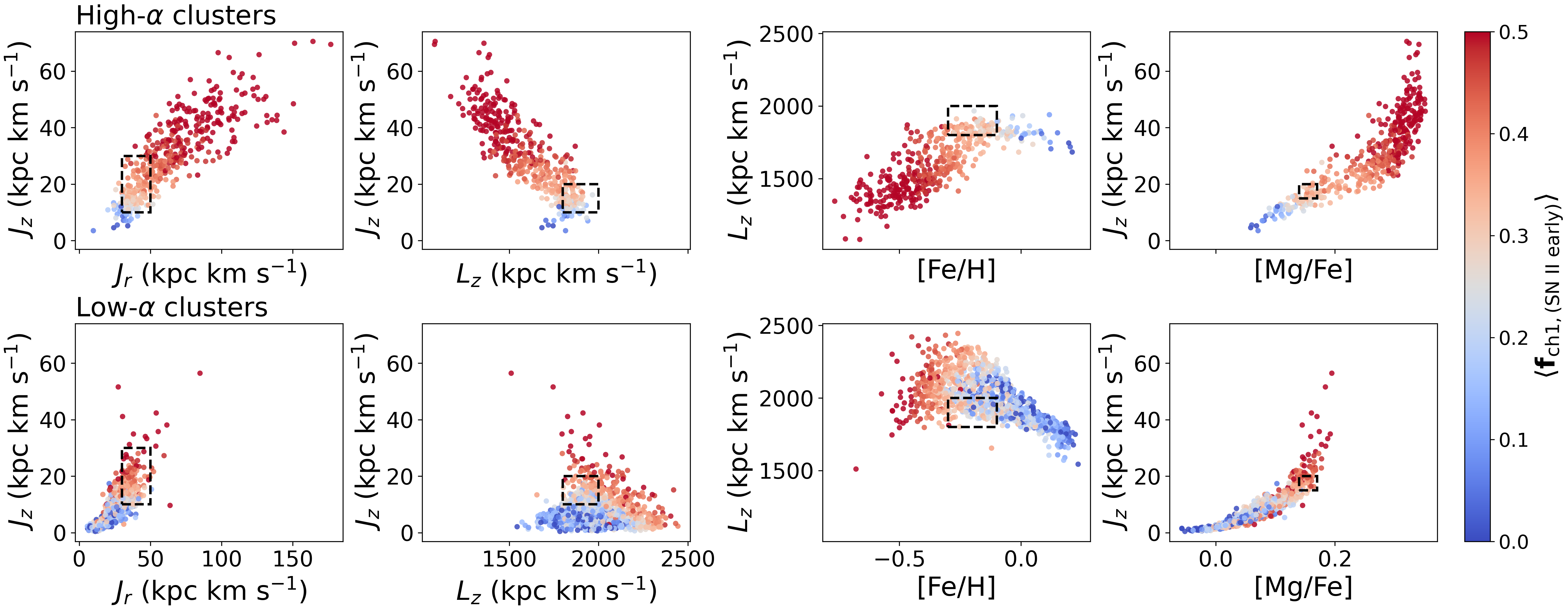}
    \caption{Distribution of latent clusters separated into high-$\alpha$ (top row) and low-$\alpha$ (bottom row) populations, projected into orbital action space and chemical planes. From left to right, panels show $J_z$ versus $J_r$, $J_z$ versus $L_z$, $J_z$ versus [Fe/H], and $J_z$ versus [Mg/Fe]. Points are coloured by the fractional contribution of the early SN~II–dominated enrichment channel, $f_{\mathrm{ch1}}$. The  high- and low-$\alpha$ discs show overlap  in orbital actions, vertical structure, and metallicity. In regions of overlap, marked in the boxes, clusters exhibit similar latent chemical fractions, indicating that the underlying enrichment varies smoothly across the disc and that the $\alpha$-bimodality reflects how a continuous enrichment sequence is expressed in specific abundance projections rather than distinct chemical populations.}
    \label{fig:alphatest}
\end{figure*}

The high-$\alpha$ and low-$\alpha$ populations shown in the top and bottom of Figure \ref{fig:alphatest} show overlap in regions of action space, vertical structure, and the metallicity--action plane. In these overlapping regions, the latent fractions, shown for channel 1 (SN~II) are indistinguishable: clusters with similar orbital and spatial properties exhibit the same relative contributions from the enrichment channels. This indicates that the chemical enrichment varies smoothly with spatial and dynamical structure, rather than being partitioned into discrete high-$\alpha$ and low-$\alpha$ regimes.  While a pronounced valley exists in projections involving $[\mathrm{Mg/Fe}]$, this reflects how a smoothly varying enrichment sequence is sampled along a particular abundance dimension, rather than a bifurcation in the underlying chemical enrichment structure of the disc.

\section{Discussion}
\label{discussion}

This study examines how the fractional contributions of shared abundance patterns relates to the chemical and spatial structure of the Milky Way. The results reveal how similar abundance patterns, rather than the metallicity scale, can capture signatures of the disc’s formation and evolution. Our approach connects well to mono-abundance population analyses \citep{Bovy2012, Bovy2016b, Mackereth2017}, but we leverage the decomposition of the full set of abundances into shared patterns that each vary per star. These per-star pattern variations serve as our precision nucleosynthesis metrics, and subsequent tags of similar enrichment environment.

We find that the relative contributions of the shared latent patterns of disc stars varies in a highly organised way across time, spatial dimensions, and metallicity.  By using latent-fraction groups, we identify empirical evolutionary ``paths''. That is, this shows an organised evolution of stars with similar abundance patterns with disc radius, height, and stellar age. These paths reveal how the vertical and radial stratification of the disc, which is set by the Galaxy’s gravitational potential, shapes the distribution of enrichment channels \citep[e.g.,][]{LSM2024}. Key transitions and perturbations imprint themselves on these trajectories, providing a new way to identify dynamical and chemical landmarks in the disc’s evolution. Our main findings are summarized as follows:

\begin{itemize}

    \item The four enrichment channels exhibit systematic changes in contribution fraction among stars, across both Galactic radius and height from the plane. Across the well-sampled region of $R_{\mbox{gal}} = 4$–$12$~kpc, the radial behaviour of pattern fractions can be described by a flexible but constrained model consisting of a shared set of five damped radial modes. This demonstrates that, despite their distinct patterns, and associated nucleosynthetic origins, the enrichment channels do not vary independently on large scales: populations with different fractional compositions are organised by the same underlying radial structure, rather than forming radially isolated sequences.

\item The age dependence of the mean latent fractions, together with increased scatter in [Fe/H] and [Mg/Fe] among stars grouped by NMF fractions, reveals a shift in disc enrichment behaviour at $\sim$6~Gyr. At this epoch, the contrast between the SN~II and SN~Ia channels decreases, and stars sharing similar
latent enrichment patterns span a broader range of absolute abundances. This indicates that similar enrichment patterns are realised across a wider range of absolute abundances, consistent with a change in how enrichment processes are sampled over time. We also observe a flattening in the valley in the [Mg/Fe] distribution when examined in terms of the difference between these channel fractions, which may reflect changes in the relative timing of enrichment contributions.

     \item Groups of stars, clustered by their NMF latent fractions, show dynamical coherence and a coupling to fractional channel contribution. Almost half are mono-age groups, with the remainder having a median age dispersion of $\approx~0.9$Gyr that serve as tracers of formation and subsequent evolution across the disc in their behaviour. The groups show a tight correlation between their mean radial and vertical actions, which clearly captures the joint radial and vertical disc heating, or relationship set at birth.

    \item The trajectory of the stars in clustered groups shows an organised and structured sequence across spatial and orbital extent. The running means of stars in each group across $R_{\mathrm{gal}}$–[Fe/H] show that, as expected, stars with similar abundance patterns span a large radial range. Similarly, running means of stars in each group across the eccentricity–$L_z$ plane shows the coherent relationship between radial location and orbital behaviour captured by the groups and their evolution.
    
 \item The running means of the NMF-defined groups in the $R_{\mathrm{gal}}$–$|z|$ plane show a systematic variation in vertical extent with Galactocentric radius, including minima in their mean height, $|z|$, at $R_{\mathrm{gal}} \sim$6~kpc and $\sim$10~kpc. These features reflect organised radial trends in the spatial distribution of chemically similar stellar populations.
 
\end{itemize}

We highlight that our analysis is limited by the 16 elemental abundances available in the sample. The inclusion of additional elements, particularly neutron-capture elements tracing s- and r-process enrichment, may reveal further enrichment pathways, and modify the detailed form of the latent patterns \citep[see][]{Ness2026}. However, the primary conclusions here should be robust across surveys.

\subsection{Implications of Chemo-Dynamical Correlations}

We find a tight coupling between the evolving spatial distribution of the enrichment patterns, and the chemical and dynamical properties of the disc.  Figures \ref{fig:spatial_channels} -- \ref{fig:latent_groups} demonstrate how enrichment modes change over time and spatial location in an organised way. For example, stars close to the plane, and younger stars, are dominated by a channel 3 (an SN~Ia pattern)  in the inner Galaxy and channel 4 (an AGB pattern) in the outer Galaxy. Conversely, at larger heights, stars at smaller radii and larger height from the plane show a high channel 1 and 2 fraction (both associated with SN II enrichment). Our NMF–clustered groups directly reveal dynamical coherence, particularly in their vertical and radial actions. While this could be interpreted as a coupling between radial and vertical heating, comparison with \citet{Frankel2020}, who argue that radial migration produces little vertical heating, indicates that stars were likely \emph{born} with coupled radial and vertical actions. These correlations reflect shared birth environments rather than migration-induced heating. Such coupling may arise naturally in an early, turbulent phase of disc evolution \citep[e.g.,][]{BH2025}, with subsequent merger-driven heating contributing additional structure. 

Complementing these insights, the running means of stars in latent groups together show radius-dependent vertical flaring for younger stars beyond $R_{\mathrm{gal}}\gtrsim8$\,kpc, and an age- and enrichment-dependent flaring pattern that agrees with previous studies \citep{Mackereth2017,CG2024,Lian2022}. In addition,  at radii near $\sim$6\,kpc and $\sim$10\,kpc, we observe local minima in the mean absolute vertical height $|z|$ (Figure~\ref{fig:disc_flaring}, at right). These features may indicate perturbations in the disc, possibly related to past dynamical disturbances.

The ensemble correlations between stellar enrichment fractions and orbital and spatial properties, together with the temporal and spatial organisation of the NMF groups, provides evidence for an interconnected disc formation history across its full extent, rather than one governed by independent or isolated processes.  Groups of stars, clustered using the NMF latent fractions, and therefore, abundance pattern similarity, span organised sequences across metallicity, orbital parameters and age. The subsequent evolutionary paths, which together these groups reveal, connect the traditionally defined low- and high-$\alpha$ discs. This is compatible with the view of the Galactic disc as a chemically coherent structure rather than two disconnected components, even though its formation history includes discrete transitions in enrichment signatures \citep[e.g.,][]{Sharma2021, Mackereth2017, Bovy2012, Bovy2016b, Schonrich2009, Haywood2008}.

Finally, we note that this analysis does not perform strong chemical tagging of individual co-natal clusters, but instead is a test and subsequent reporting of the large-scale correlated enrichment structure and dynamical assembly history of the Galactic disc.

\subsection{Interpreting the $\alpha$-Bimodality in a Chemically Continuous disc}

Previous studies have examined the [Fe/H]-[$\alpha$/Fe] relation across different Galactic components, or across stellar ages \citep[e.g.,][]{Adi2012, Bovy2012, Bensby2014, Lu2022a, Gandhi2019, Nidever2014, Hayden2015, Mackereth2017, Miglio2021, Queiroz2023, Sharma2021, Haywood2013, Belokurov2022, SilvaAguirre2018, Ness2019, Lu2022a, Ratcliffe2024, Lu2022}. These studies find that the chemical and orbital properties of the disc are closely linked and mono-abundance or mono-age populations that are structured and organised spatially reflect this connection. Our work extends that framework with a fully data-driven approach using abundance pattern similarity; although we associate the different latent patterns uncovered in Paper~I with discrete sources, our conclusions would remain intact if we did not.  

The spatial and orbital coherence of mono-age and mono-abundance populations \citep[e.g.][]{Mackereth2017, Bovy2012}, juxtaposed with the bimodality in the [Fe/H]–[$\alpha$/Fe] plane has led to a longstanding conceptual tension. This has motivated the idea that the high- and low-$\alpha$ sequences trace two fundamentally separate stellar populations. Our analysis indicates the Milky Way disc is a coherent structure and the full population can be described by shared patterns, with a changing contributions of these patterns.

While we interpret the high- and low-$\alpha$ sequences as two phases of a connected enrichment trajectory, our results do not exclude mergers or external perturbations that alter the star formation rate or gas inflow. Rather, they constrain any such events, to the extent that they are reflected in stellar abundances, to remain chemically coupled to the disc’s enrichment history, rather than introducing chemically autonomous populations. One possibility is that nucleosynthetic yields are sufficiently universal that diverse formation environments project onto a common low-dimensional abundance space. In this case, the absence of chemically autonomous populations would reflect a physical property of chemical enrichment rather than a limitation of the model. This hypothesis can be tested through controlled comparisons with simulations that include chemically distinct accretion events. This can also be tested empirically by applying the same latent enrichment framework to dwarf galaxies, whose bursty star formation histories, stochastic enrichment, and inefficient mixing \citep{Tolstoy2009} provide a stringent test of chemical universality.

We identify the transition epoch in the enrichment mode at $\approx$6~Gyr.  We interpret this change as reflecting the change in enrichment phase. This epoch lies close to reported passages of the Sagittarius dwarf galaxy \citep[e.g.][]{RL2020,
Laporte2018, Lu2024}, and it is possible that external perturbations may have
contributed to the observed transition. We also note the highest dispersion in [Fe/H] and [Mg/Fe] of the groups at $\approx$6-7 Gyr; coincident with the largest spread in the age-metallicity relation seen at a similar epoch seen using Gaia \citep{FA2025}. The transition at $\sim$6~Gyr and highest intrinsic scatter in the groups at 6-7~Gyr also aligns with features reported in multiple independent analyses. \citet{VC2025} find evidence for gas dilution at 7--9~Gyr in \apogee\ data. The birth-radius analysis of \citet{Ratcliffe2024} identifies changes near $\sim$6~Gyr (and additional signatures around 9~Gyr), potentially corresponding to the spatial perturbations we observe in our oldest groups (Figure~\ref{fig:disc_flaring}). Using \apogee{}, \citet{Katz2021} also infer renewed gas inflow at $\sim$6~Gyr, immediately preceding the emergence of the low-$\alpha$ disc.

A wide range of models attribute the $\alpha$-bimodality to changes in the Galaxy’s gas accretion or star-formation history. These include the classical two-infall model \citep{Chiappini1997, Vincenzo2020, Grand2020, Spitoni2019} and its extensions, such as the three-infall model of \citet{Spitoni2023}, as well as models that incorporate quenching phases or pauses in star formation followed by renewed gas accretion \citep[e.g.][]{Haywood2013, Snaith2015, Snaith2022, Haywood2016, Haywood2015, Haywood2018, Katz2021, Lian2020a, Lian2020, Spitoni2019, Spitoni2021, Spitoni2022, ES2024}. Other studies have demonstrated that the $\alpha$-bimodality can arise without discrete accretion episodes, instead invoking smooth transitions in accretion mode or the effects of radial migration \citep{Chen2023, Sharma2021}. Our results do not distinguish between specific implementations of gas accretion or star-formation history (e.g. two-infall, quenching episodes, or smoothly varying accretion). Instead, they place empirical constraints on the class of chemical evolution scenarios that are consistent with the observed abundance structure of the Milky Way disc. The observations favor models in which star formation proceeds through system-dependent trajectories within a common, low-dimensional enrichment framework.

The equilibrium models of \citet{Johnson2025} demonstrate that the Galaxy can maintain a steady balance between inflows and outflows that regulates its chemical evolution. Our results are congruent with this picture, in which regulated gas inflow, modulated by the Galaxy’s mass, underpins the disc’s ordered chemical structure \citep[][]{Bird2021, Aumer2016}. This is provided that a mechanism exists to drive the observed shift in the dominant enrichment source, which we do not test directly. The tight coupling we observe between orbital, spatial, and chemical properties suggests that self-regulation is a plausible framework, in which changes in enrichment reflect disc settling within the Galaxy’s gravitational potential. This interpretation is further supported by chemo-dynamical simulations showing that merger-driven gas infall is not required to generate the $\alpha$-bimodality \citep{Kh2021}. Our results are also compatible with \citet{lb2025}, who argue for an internally driven origin of the low-$\alpha$ disc, although we do not find evidence supporting a merger-induced truncation of high-$\alpha$ star formation as proposed in that work. Mergers or external gas accretion are not excluded by our results; however, any such events must be chemically consistent with the Galaxy’s ongoing enrichment history, rather than introducing chemically autonomous stellar populations.

The continuous yet evolving enrichment behaviour we identify is consistent with analytic and semi-analytic models that describe the Milky Way disc as a unified, evolving system \citep[e.g.][]{Schonrich2009, Haywood2013, Sharma2022, Bovy2016b, Minchev2013, Kh2021, Johnson2025, Chen2023}, as well as simulations in which the low-$\alpha$ sequence emerges from renewed accretion of metal-poor circumgalactic gas \citep{Hanna2025, Ork2025}. In these frameworks, the observed abundance structure arises without the need for sharp inflow discontinuities, instead reflecting gradually evolving inflow conditions regulated by the Galaxy’s mass. Within this broader picture, the thick (high-$\alpha$) disc forms during an early phase of turbulent accretion and baryon “sloshing” \citep{BH2025}, followed by a more quiescent, inflow-regulated phase that builds the thin (low-$\alpha$) disc. While more detailed, model-specific comparisons are required to discriminate among these scenarios, the smoothly varying and highly ordered enrichment paths we observe indicate that chemical evolution proceeded largely continuously, with changes primarily in rate and spatial distribution rather than in the underlying enrichment structure.

\section{Conclusion}
\label{conclusion}

Disentangling stellar abundance information through NMF decomposition, clustering, and analysis of inter- and intra-group behaviour shows the Milky Way disc is a strongly coupled chemo-dynamical system across its full extent. We find correlations between latent abundance fractions and stellar age, orbital properties, and spatial structure. This shows that stars sharing similar enrichment patterns also share coherent dynamical and temporal characteristics.

The latent enrichment fractions define a continuous path across radius, height, and age, rather than isolated or disconnected populations. Transitions in the dominant enrichment contributions mark changes in the relative importance of enrichment patterns (and therefore their associated nucleosynthetic sources), but these occur within a shared low-dimensional abundance structure. The smooth variation of enrichment fractions across spatial and dynamical variables demonstrates that disc chemical evolution remains organised and structured, even as enrichment rates and source contributions change over time. While external processes such as mergers or gas accretion are not ruled out, our findings mean that any such events must preserve continuity in abundance patterns and their spatial and orbital correlations, rather than introducing chemically autonomous stellar populations.

\section*{Data Availability}

The data used here are part of the SDSS-V Milky Way Mapper survey, and are publicly available through the Sloan Digital Sky Survey data releases (DR19). 

\section{Acknowledgements}

Funding for the Sloan Digital Sky Survey V has been provided by the Alfred P. Sloan Foundation, the Heising-Simons Foundation, the National Science Foundation, and the Participating Institutions. SDSS acknowledges support and resources from the Center for High-Performance Computing at the University of Utah. SDSS telescopes are located at Apache Point Observatory, funded by the Astrophysical Research Consortium and operated by New Mexico State University, and at Las Campanas Observatory, operated by the Carnegie Institution for Science. The SDSS web site is \url{www.sdss.org}.

SDSS is managed by the Astrophysical Research Consortium for the Participating Institutions of the SDSS Collaboration, including the Carnegie Institution for Science, Chilean National Time Allocation Committee (CNTAC) ratified researchers, Caltech, the Gotham Participation Group, Harvard University, Heidelberg University, The Flatiron Institute, The Johns Hopkins University, L'Ecole polytechnique f\'{e}d\'{e}rale de Lausanne (EPFL), Leibniz-Institut f\"{u}r Astrophysik Potsdam (AIP), Max-Planck-Institut f\"{u}r Astronomie (MPIA Heidelberg), Max-Planck-Institut f\"{u}r Extraterrestrische Physik (MPE), Nanjing University, National Astronomical Observatories of China (NAOC), New Mexico State University, The Ohio State University, Pennsylvania State University, Smithsonian Astrophysical Observatory, Space Telescope Science Institute (STScI), the Stellar Astrophysics Participation Group, Universidad Nacional Aut\'{o}noma de M\'{e}xico, University of Arizona, University of Colorado Boulder, University of Illinois at Urbana-Champaign, University of Toronto, University of Utah, University of Virginia, Yale University, and Yunnan University.

R. L-V. acknowledges support from Secretar\'ia de Ciencia, Humanidades, Tecnolog\'ia e Inovacci\'on (SECIHTI) through a postdoctoral fellowship within the program ``Estancias posdoctorales por M\'exico''

LC thanks the support from grant IA103326 (DGAPA-PAPIIT, UNAM) 

Some text and code refinements were developed in a highly supervised engagement with AI language models  (ChatGPT, OpenAI and Gemini, Google).

\section{Appendix}

\subsection{
Model Patterns and Validation}

\begin{figure}
    \centering
    \includegraphics[width=1\linewidth]{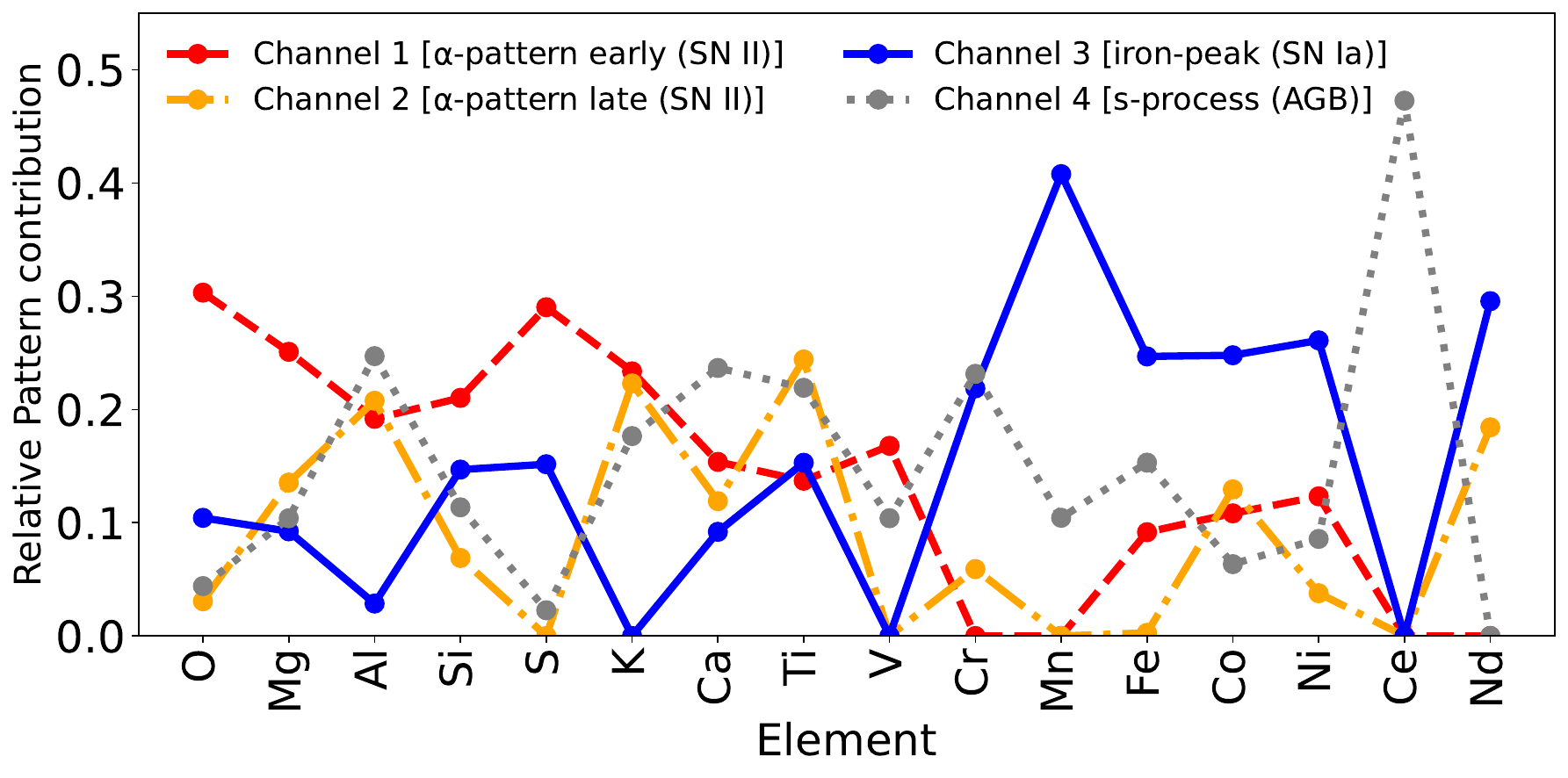}
\caption{The four latent pattern vectors, $P$, identified from data, showing the relative contribution of each element to each pattern. These patterns do not represent single yields, but rather integrated enrichment signatures. They each show distinct behaviour, and can subsequently be associated with dominant enrichment channels, or sources, as indicated in the legend. There are some differences in these pattern vectors compared to Paper I but they show the same overall behaviour and we make the same subsequent association with sources. Channel~1: Early SN~II–dominated enrichment from massive core-collapse supernovae, characterised by high contributions of O, Mg, S, and Si.  Channel~2: Late SN~II–dominated enrichment associated with lower-mass core-collapse supernovae, producing enhanced Si, Ca, Ti, and K, with secondary contributions to Al. Channel~3: Iron-peak enrichment pattern associated with thermonuclear SN~Ia, contributing Cr, Mn, Fe, Co, and Ni on longer timescales.  Channel~4: Delayed enrichment, dominated by the s-process with a pronounced Ce peak, arising primarily from asymptotic giant branch (AGB) stars returning heavy elements through stellar winds, together with correlated structure in K, Ca, Ti, and Cr, likely reflecting metallicity-dependent yields rather than direct AGB production. While we associate each channel with characteristic nucleosynthetic sources, these components are not pure tracers of individual processes and are formally mathematical representations of covariance among elemental abundances. The physical interpretation of each channel is tested using independent information, such as stellar ages, spatial distributions, and orbital properties.}
    \label{fig:appendix1}
\end{figure}

In this paper we employ the same framework as in Paper I (Ness et al., 2026, submitted) but on a larger set of stars across the disc. In this work we run the NMF decomposition on a parent set of 199,290 stars compared to  70,057 in Paper I. We select for the same number of latent variables ($M=4$) which well describe the sample with a reduced $\chi^2$ mode of 1.2. We show in Figure \ref{fig:appendix1} the pattern decomposition of the abundances. 
The NMF decomposition returns a set of latent abundance patterns together with coefficients describing the contribution of each pattern to each star. The quantities shown are the direct amplitudes of the latent patterns returned by the NMF decomposition and are not independently normalised. The physically relevant information is contained in the relative structure of the element amplitudes within each latent channel, and in the relative contribution of the channels between stars.

\begin{figure}
    \centering
    \includegraphics[width=1\linewidth]{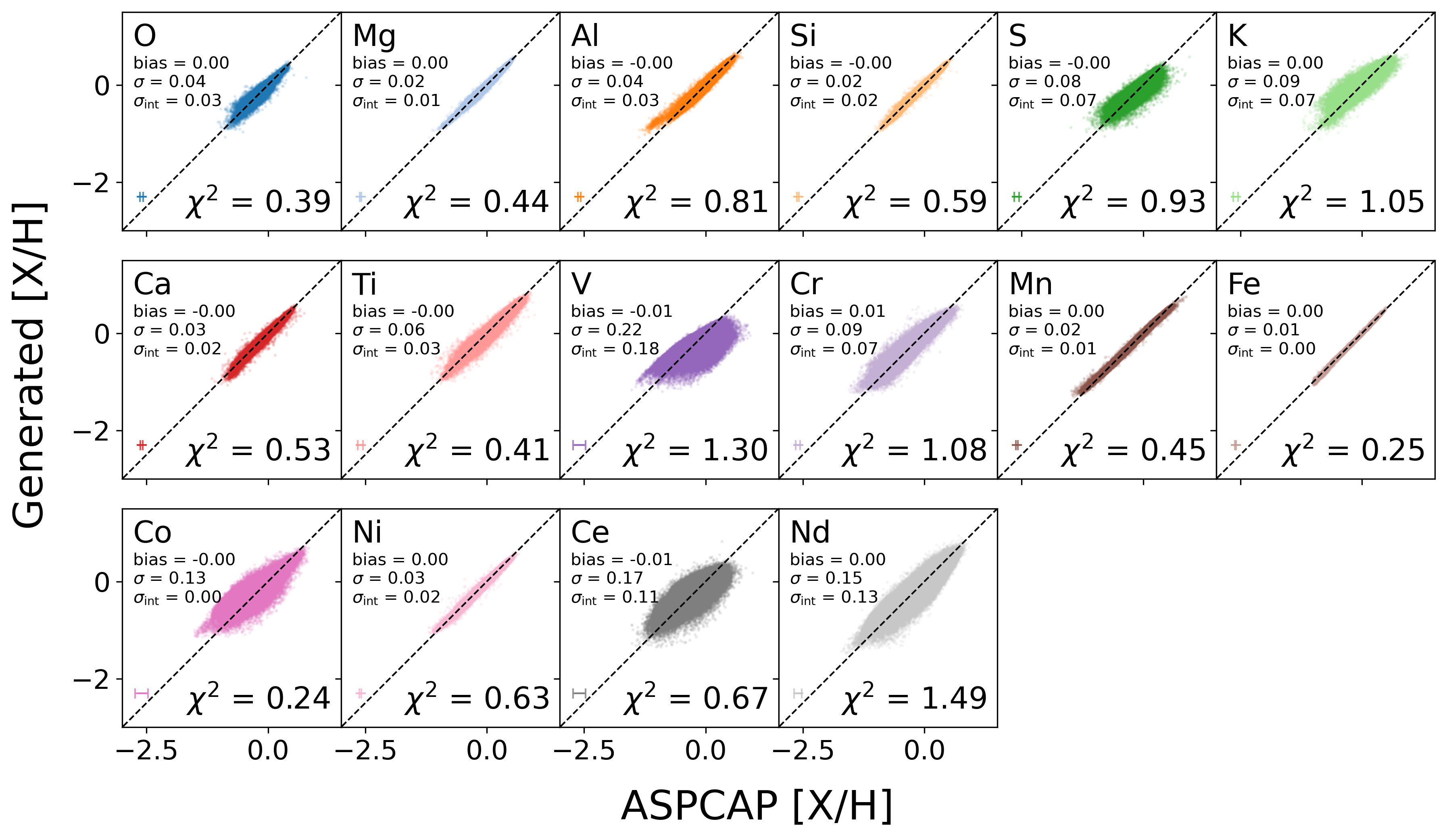}
    \caption{Elemental abundances generated by the latent variable model for the Galactic Genesis disc sample, using $M = 4$ latent components. The measured \textsc{ASTRA ASPCAP}  abundances are on the x-axis and the y-axis reports the generated abundance using the NMF basis. The bias, scatter and intrinsic scatter (subtracting the mean error in quadrature) are indicated in each sub-figure.}
    \label{fig:appendix2}
\end{figure}

Figure \ref{fig:appendix2} shows the measured \textsc{ASTRA ASPCAP} abundances compared to the abundances generated from the NMF latent basis with $M=4$.

\subsection{Residuals in the disc Plane Away from Radial Model}

\begin{figure}
    \centering
    \includegraphics[width=1.\linewidth]{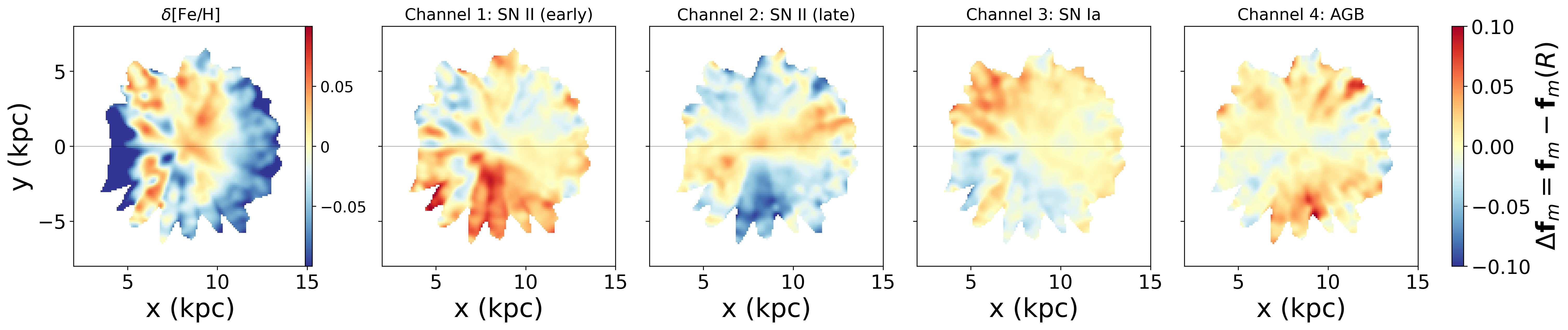}
    \caption{Residual structure in the Galactic $x$--$y$ plane after subtraction of smooth radial trends. The left panel shows $\delta[\mathrm{Fe/H}]$, defined as the residual after subtracting a mean radial [Fe/H] gradient. The four right panels show residuals in the NMF channel fractions, $\Delta f_m = f_m - f_m(R)$, after subtraction of the shared radial mode model shown in Figure~\ref{fig:damped}. Positive values indicate spatial regions where a given enrichment channel is over-represented relative to the mean at that Galactocentric radius, while negative values indicate under-representation. The channel residuals reveal low-amplitude azimuthal inhomogeneities that are not similar to the [Fe/H] residual field.}
    \label{fig:residuals}
\end{figure}

In Section~3.1.1, we show that the radial behaviour of the four latent enrichment channels can be fairly well described by a shared radial mode model. We examine residuals about this model in Figure~\ref{fig:residuals}, which shows azimuthal inhomogeneities. The left panel shows the Galactic $x$--$y$ map of residual [Fe/H] after subtraction of the mean radial [Fe/H] gradient; this shows structure similar to that previously reported by \citet{Hawkshaw2024}. The four panels on the right show the corresponding residuals in the four channel fractions for stars with $|z|< 3.0$~kpc, using the radial model shown in Figure~\ref{fig:damped}.  The panel comparisons show that the enrichment channel residuals show inhomogeneities that are distinct from the [Fe/H] residual field, but are of comparable amplitude. These enrichment channel fraction residuals may be inherited from local under- or over-representations of each enrichment channel relative to the mean contribution at that radius. A detailed interpretation of this structure will require larger and more complete Milky Way Mapper samples, particularly from the Galactic Genesis program which aims to obtain stars across the Milky Way disc for $\approx 3$ million stars \citep{Kollmeier2025}.

\begin{thebibliography}{}
\makeatletter
\relax
\def\mn@urlcharsother{\let\do\@makeother \do\$\do\&\do\#\do\^\do\_\do\%\do\~}
\def\mn@doi{\begingroup\mn@urlcharsother \@ifnextchar [ {\mn@doi@}
  {\mn@doi@[]}}
\def\mn@doi@[#1]#2{\def\@tempa{#1}\ifx\@tempa\@empty \href
  {http://dx.doi.org/#2} {doi:#2}\else \href {http://dx.doi.org/#2} {#1}\fi
  \endgroup}
\def\mn@eprint#1#2{\mn@eprint@#1:#2::\@nil}
\def\mn@eprint@arXiv#1{\href {http://arxiv.org/abs/#1} {{\tt arXiv:#1}}}
\def\mn@eprint@dblp#1{\href {http://dblp.uni-trier.de/rec/bibtex/#1.xml}
  {dblp:#1}}
\def\mn@eprint@#1:#2:#3:#4\@nil{\def\@tempa {#1}\def\@tempb {#2}\def\@tempc
  {#3}\ifx \@tempc \@empty \let \@tempc \@tempb \let \@tempb \@tempa \fi \ifx
  \@tempb \@empty \def\@tempb {arXiv}\fi \@ifundefined
  {mn@eprint@\@tempb}{\@tempb:\@tempc}{\expandafter \expandafter \csname
  mn@eprint@\@tempb\endcsname \expandafter{\@tempc}}}

\bibitem[\protect\citeauthoryear{{Adibekyan}, {Sousa}, {Santos}, {Delgado
  Mena}, {Gonz{\'a}lez Hern{\'a}ndez}, {Israelian}, {Mayor}  \&
  {Khachatryan}}{{Adibekyan} et~al.}{2012}]{Adi2012}
{Adibekyan} V.~Z.,  {Sousa} S.~G.,  {Santos} N.~C.,  {Delgado Mena} E.,
  {Gonz{\'a}lez Hern{\'a}ndez} J.~I.,  {Israelian} G.,  {Mayor} M.,
  {Khachatryan} G.,  2012, \mn@doi [\aap] {10.1051/0004-6361/201219401}, \href
  {https://ui.adsabs.harvard.edu/abs/2012A&A...545A..32A} {545, A32}

\bibitem[\protect\citeauthoryear{{Adibekyan} et~al.,}{{Adibekyan}
  et~al.}{2013}]{Adi2013}
{Adibekyan} V.~Z.,  et~al., 2013, \mn@doi [\aap] {10.1051/0004-6361/201321520},
  \href {http://adsabs.harvard.edu/abs/2013A%26A...554A..44A} {554, A44}

\bibitem[\protect\citeauthoryear{{Agertz} et~al.,}{{Agertz}
  et~al.}{2021}]{Agertz2021}
{Agertz} O.,  et~al., 2021, \mn@doi [\mnras] {10.1093/mnras/stab322}, \href
  {https://ui.adsabs.harvard.edu/abs/2021MNRAS.503.5826A} {503, 5826}

\bibitem[\protect\citeauthoryear{{Anders} et~al.,}{{Anders}
  et~al.}{2023}]{Anders2023}
{Anders} F.,  et~al., 2023, \mn@doi [\aap] {10.1051/0004-6361/202346666}, \href
  {https://ui.adsabs.harvard.edu/abs/2023A&A...678A.158A} {678, A158}

\bibitem[\protect\citeauthoryear{{Antoja} et~al.,}{{Antoja}
  et~al.}{2018}]{Antoja2018}
{Antoja} T.,  et~al., 2018, \mn@doi [\nat] {10.1038/s41586-018-0510-7}, \href
  {http://adsabs.harvard.edu/abs/2018Natur.561..360A} {561, 360}

\bibitem[\protect\citeauthoryear{{Aumer}, {Binney}  \& {Sch{\"o}nrich}}{{Aumer}
  et~al.}{2016}]{Aumer2016}
{Aumer} M.,  {Binney} J.,   {Sch{\"o}nrich} R.,  2016, \mn@doi [\mnras]
  {10.1093/mnras/stw777}, \href
  {https://ui.adsabs.harvard.edu/abs/2016MNRAS.459.3326A} {459, 3326}

\bibitem[\protect\citeauthoryear{{Bailer-Jones}, {Rybizki}, {Fouesneau},
  {Demleitner}  \& {Andrae}}{{Bailer-Jones} et~al.}{2021}]{BailerJones2021}
{Bailer-Jones} C.~A.~L.,  {Rybizki} J.,  {Fouesneau} M.,  {Demleitner} M.,
  {Andrae} R.,  2021, \mn@doi [\aj] {10.3847/1538-3881/abd806}, \href
  {https://ui.adsabs.harvard.edu/abs/2021AJ....161..147B} {161, 147}

\bibitem[\protect\citeauthoryear{{Beane}, {Ness}  \& {Bedell}}{{Beane}
  et~al.}{2018}]{Beane2018}
{Beane} A.,  {Ness} M.~K.,   {Bedell} M.,  2018, \mn@doi [\apj]
  {10.3847/1538-4357/aae07f}, \href
  {https://ui.adsabs.harvard.edu/abs/2018ApJ...867...31B} {867, 31}

\bibitem[\protect\citeauthoryear{{Belokurov} \& {Evans}}{{Belokurov} \&
  {Evans}}{2022}]{Belokurov2022}
{Belokurov} V.,  {Evans} N.~W.,  2022, \mn@doi [Nature Astronomy]
  {10.1038/s41550-022-01740-w}, \href
  {https://ui.adsabs.harvard.edu/abs/2022NatAs...6..911B} {6, 911}

\bibitem[\protect\citeauthoryear{{Belokurov}, {Sanders}, {Fattahi}, {Smith},
  {Deason}, {Evans}  \& {Grand}}{{Belokurov} et~al.}{2020}]{Belokurovsplash}
{Belokurov} V.,  {Sanders} J.~L.,  {Fattahi} A.,  {Smith} M.~C.,  {Deason}
  A.~J.,  {Evans} N.~W.,   {Grand} R. J.~J.,  2020, \mn@doi [\mnras]
  {10.1093/mnras/staa876}, \href
  {https://ui.adsabs.harvard.edu/abs/2020MNRAS.494.3880B} {494, 3880}

\bibitem[\protect\citeauthoryear{{Bensby}, {Feltzing}  \&
  {Lundstr{\"o}m}}{{Bensby} et~al.}{2003}]{Bensby2003}
{Bensby} T.,  {Feltzing} S.,   {Lundstr{\"o}m} I.,  2003, \mn@doi [\aap]
  {10.1051/0004-6361:20031213}, \href
  {https://ui.adsabs.harvard.edu/abs/2003A&A...410..527B} {410, 527}

\bibitem[\protect\citeauthoryear{{Bensby}, {Feltzing}  \& {Oey}}{{Bensby}
  et~al.}{2014}]{Bensby2014}
{Bensby} T.,  {Feltzing} S.,   {Oey} M.~S.,  2014, \mn@doi [\aap]
  {10.1051/0004-6361/201322631}, \href
  {https://ui.adsabs.harvard.edu/abs/2014A&A...562A..71B} {562, A71}

\bibitem[\protect\citeauthoryear{{Bird}, {Loebman}, {Weinberg}, {Brooks},
  {Quinn}  \& {Christensen}}{{Bird} et~al.}{2021}]{Bird2021}
{Bird} J.~C.,  {Loebman} S.~R.,  {Weinberg} D.~H.,  {Brooks} A.~M.,  {Quinn}
  T.~R.,   {Christensen} C.~R.,  2021, \mn@doi [\mnras]
  {10.1093/mnras/stab289}, \href
  {https://ui.adsabs.harvard.edu/abs/2021MNRAS.503.1815B} {503, 1815}

\bibitem[\protect\citeauthoryear{{Bland-Hawthorn}, {Krumholz}  \&
  {Freeman}}{{Bland-Hawthorn} et~al.}{2010}]{BH2010}
{Bland-Hawthorn} J.,  {Krumholz} M.~R.,   {Freeman} K.,  2010, \mn@doi [\apj]
  {10.1088/0004-637X/713/1/166}, \href
  {http://adsabs.harvard.edu/abs/2010ApJ...713..166B} {713, 166}

\bibitem[\protect\citeauthoryear{{Bland-Hawthorn} et~al.,}{{Bland-Hawthorn}
  et~al.}{2025}]{BH2025}
{Bland-Hawthorn} J.,  et~al., 2025, \mn@doi [arXiv e-prints]
  {10.48550/arXiv.2502.01895}, \href
  {https://ui.adsabs.harvard.edu/abs/2025arXiv250201895B} {p. arXiv:2502.01895}

\bibitem[\protect\citeauthoryear{{Bonaca} et~al.,}{{Bonaca}
  et~al.}{2020}]{Bonaca2020}
{Bonaca} A.,  et~al., 2020, \mn@doi [\apjl] {10.3847/2041-8213/ab9caa}, \href
  {https://ui.adsabs.harvard.edu/abs/2020ApJ...897L..18B} {897, L18}

\bibitem[\protect\citeauthoryear{{Borbolato} et~al.,}{{Borbolato}
  et~al.}{2025}]{lb2025}
{Borbolato} L.,  et~al., 2025, \mn@doi [arXiv e-prints]
  {10.48550/arXiv.2504.00135}, \href
  {https://ui.adsabs.harvard.edu/abs/2025arXiv250400135B} {p. arXiv:2504.00135}

\bibitem[\protect\citeauthoryear{{Bovy}}{{Bovy}}{2015}]{galpy}
{Bovy} J.,  2015, \mn@doi [\apjs] {10.1088/0067-0049/216/2/29}, \href
  {https://ui.adsabs.harvard.edu/abs/2015ApJS..216...29B} {216, 29}

\bibitem[\protect\citeauthoryear{{Bovy}, {Rix}, {Liu}, {Hogg}, {Beers}  \&
  {Lee}}{{Bovy} et~al.}{2012a}]{Bovy2012}
{Bovy} J.,  {Rix} H.-W.,  {Liu} C.,  {Hogg} D.~W.,  {Beers} T.~C.,   {Lee}
  Y.~S.,  2012a, \mn@doi [\apj] {10.1088/0004-637X/753/2/148}, \href
  {https://ui.adsabs.harvard.edu/abs/2012ApJ...753..148B} {753, 148}

\bibitem[\protect\citeauthoryear{{Bovy}, {Rix}, {Hogg}, {Beers}, {Lee}  \&
  {Zhang}}{{Bovy} et~al.}{2012b}]{Bovy2012b}
{Bovy} J.,  {Rix} H.-W.,  {Hogg} D.~W.,  {Beers} T.~C.,  {Lee} Y.~S.,   {Zhang}
  L.,  2012b, \mn@doi [\apj] {10.1088/0004-637X/755/2/115}, \href
  {https://ui.adsabs.harvard.edu/abs/2012ApJ...755..115B} {755, 115}

\bibitem[\protect\citeauthoryear{{Bovy}, {Rix}, {Schlafly}, {Nidever},
  {Holtzman}, {Shetrone}  \& {Beers}}{{Bovy} et~al.}{2016}]{Bovy2016b}
{Bovy} J.,  {Rix} H.-W.,  {Schlafly} E.~F.,  {Nidever} D.~L.,  {Holtzman}
  J.~A.,  {Shetrone} M.,   {Beers} T.~C.,  2016, \mn@doi [\apj]
  {10.3847/0004-637X/823/1/30}, \href
  {https://ui.adsabs.harvard.edu/abs/2016ApJ...823...30B} {823, 30}

\bibitem[\protect\citeauthoryear{{Bowen} \& {Vaughan}}{{Bowen} \&
  {Vaughan}}{1973}]{Bowen1973}
{Bowen} I.~S.,  {Vaughan} Jr. A.~H.,  1973, \mn@doi [\ao]
  {10.1364/AO.12.001430}, \href
  {https://ui.adsabs.harvard.edu/abs/1973ApOpt..12.1430B} {12, 1430}

\bibitem[\protect\citeauthoryear{{Brook} et~al.,}{{Brook}
  et~al.}{2012}]{Brook2012}
{Brook} C.~B.,  et~al., 2012, \mn@doi [\mnras]
  {10.1111/j.1365-2966.2012.21738.x}, \href
  {https://ui.adsabs.harvard.edu/abs/2012MNRAS.426..690B} {426, 690}

\bibitem[\protect\citeauthoryear{{Buck}}{{Buck}}{2020}]{Buck2020}
{Buck} T.,  2020, \mn@doi [\mnras] {10.1093/mnras/stz3289}, \href
  {https://ui.adsabs.harvard.edu/abs/2020MNRAS.491.5435B} {491, 5435}

\bibitem[\protect\citeauthoryear{{Cantat-Gaudin} et~al.,}{{Cantat-Gaudin}
  et~al.}{2024}]{CG2024}
{Cantat-Gaudin} T.,  et~al., 2024, \mn@doi [\aap]
  {10.1051/0004-6361/202348018}, \href
  {https://ui.adsabs.harvard.edu/abs/2024A&A...683A.128C} {683, A128}

\bibitem[\protect\citeauthoryear{{Casali} et~al.,}{{Casali}
  et~al.}{2023}]{Casali2023}
{Casali} G.,  et~al., 2023, \mn@doi [\aap] {10.1051/0004-6361/202346274}, \href
  {https://ui.adsabs.harvard.edu/abs/2023A&A...677A..60C} {677, A60}

\bibitem[\protect\citeauthoryear{{Cerqui}, {Haywood}, {Snaith}, {Di Matteo}  \&
  {Casamiquela}}{{Cerqui} et~al.}{2025}]{VC2025}
{Cerqui} V.,  {Haywood} M.,  {Snaith} O.,  {Di Matteo} P.,   {Casamiquela} L.,
  2025, \mn@doi [\aap] {10.1051/0004-6361/202452448}, \href
  {https://ui.adsabs.harvard.edu/abs/2025A&A...699A.277C} {699, A277}

\bibitem[\protect\citeauthoryear{{Chandra} et~al.,}{{Chandra}
  et~al.}{2023}]{Chandra2023}
{Chandra} V.,  et~al., 2023, \mn@doi [arXiv e-prints]
  {10.48550/arXiv.2310.13050}, \href
  {https://ui.adsabs.harvard.edu/abs/2023arXiv231013050C} {p. arXiv:2310.13050}

\bibitem[\protect\citeauthoryear{{Chen}, {Hayden}, {Sharma}, {Bland-Hawthorn},
  {Kobayashi}  \& {Karakas}}{{Chen} et~al.}{2023}]{Chen2023}
{Chen} B.,  {Hayden} M.~R.,  {Sharma} S.,  {Bland-Hawthorn} J.,  {Kobayashi}
  C.,   {Karakas} A.~I.,  2023, \mn@doi [\mnras] {10.1093/mnras/stad1568},
  \href {https://ui.adsabs.harvard.edu/abs/2023MNRAS.523.3791C} {523, 3791}

\bibitem[\protect\citeauthoryear{{Chiappini}, {Matteucci}  \&
  {Gratton}}{{Chiappini} et~al.}{1997}]{Chiappini1997}
{Chiappini} C.,  {Matteucci} F.,   {Gratton} R.,  1997, \mn@doi [\apj]
  {10.1086/303726}, \href
  {https://ui.adsabs.harvard.edu/abs/1997ApJ...477..765C} {477, 765}

\bibitem[\protect\citeauthoryear{{Eilers}, {Hogg}, {Rix}, {Ness},
  {Price-Whelan}, {M{\'e}sz{\'a}ros}  \& {Nitschelm}}{{Eilers}
  et~al.}{2022}]{Eilers2022}
{Eilers} A.-C.,  {Hogg} D.~W.,  {Rix} H.-W.,  {Ness} M.~K.,  {Price-Whelan}
  A.~M.,  {M{\'e}sz{\'a}ros} S.,   {Nitschelm} C.,  2022, \mn@doi [\apj]
  {10.3847/1538-4357/ac54ad}, \href
  {https://ui.adsabs.harvard.edu/abs/2022ApJ...928...23E} {928, 23}

\bibitem[\protect\citeauthoryear{{Fernandez-Alvar} et~al.,}{{Fernandez-Alvar}
  et~al.}{2025}]{FA2025}
{Fernandez-Alvar} E.,  et~al., 2025, \mn@doi [arXiv e-prints]
  {10.48550/arXiv.2503.19536}, \href
  {https://ui.adsabs.harvard.edu/abs/2025arXiv250319536F} {p. arXiv:2503.19536}

\bibitem[\protect\citeauthoryear{{Frankel}, {Rix}, {Ting}, {Ness}  \&
  {Hogg}}{{Frankel} et~al.}{2018}]{Frankel2018}
{Frankel} N.,  {Rix} H.-W.,  {Ting} Y.-S.,  {Ness} M.~K.,   {Hogg} D.~W.,
  2018, preprint, \href {http://adsabs.harvard.edu/abs/2018arXiv180509198F} {}
  (\mn@eprint {arXiv} {1805.09198})

\bibitem[\protect\citeauthoryear{{Frankel}, {Sanders}, {Rix}, {Ting}  \&
  {Ness}}{{Frankel} et~al.}{2019}]{Frankel2019}
{Frankel} N.,  {Sanders} J.,  {Rix} H.-W.,  {Ting} Y.-S.,   {Ness} M.,  2019,
  \mn@doi [\apj] {10.3847/1538-4357/ab4254}, \href
  {https://ui.adsabs.harvard.edu/abs/2019ApJ...884...99F} {884, 99}

\bibitem[\protect\citeauthoryear{{Frankel}, {Sanders}, {Ting}  \&
  {Rix}}{{Frankel} et~al.}{2020}]{Frankel2020}
{Frankel} N.,  {Sanders} J.,  {Ting} Y.-S.,   {Rix} H.-W.,  2020, \mn@doi
  [\apj] {10.3847/1538-4357/ab910c}, \href
  {https://ui.adsabs.harvard.edu/abs/2020ApJ...896...15F} {896, 15}

\bibitem[\protect\citeauthoryear{{Fuhrmann}}{{Fuhrmann}}{1998}]{Fuhrmann1998}
{Fuhrmann} K.,  1998, \aap, \href
  {https://ui.adsabs.harvard.edu/abs/1998A&A...338..161F} {338, 161}

\bibitem[\protect\citeauthoryear{{Gaia Collaboration} et~al.,}{{Gaia
  Collaboration} et~al.}{2023}]{GC1}
{Gaia Collaboration} et~al., 2023, \mn@doi [\aap]
  {10.1051/0004-6361/202243940}, \href
  {https://ui.adsabs.harvard.edu/abs/2023A&A...674A...1G} {674, A1}

\bibitem[\protect\citeauthoryear{{Gandhi} \& {Ness}}{{Gandhi} \&
  {Ness}}{2019}]{Gandhi2019}
{Gandhi} S.~S.,  {Ness} M.~K.,  2019, arXiv e-prints, \href
  {http://adsabs.harvard.edu/abs/2019arXiv190304030G} {}

\bibitem[\protect\citeauthoryear{{Gilmore} \& {Reid}}{{Gilmore} \&
  {Reid}}{1983}]{Gilmore1983}
{Gilmore} G.,  {Reid} N.,  1983, \mn@doi [\mnras] {10.1093/mnras/202.4.1025},
  \href {https://ui.adsabs.harvard.edu/abs/1983MNRAS.202.1025G} {202, 1025}

\bibitem[\protect\citeauthoryear{{Grand} et~al.,}{{Grand}
  et~al.}{2020}]{Grand2020}
{Grand} R. J.~J.,  et~al., 2020, \mn@doi [\mnras] {10.1093/mnras/staa2057},
  \href {https://ui.adsabs.harvard.edu/abs/2020MNRAS.497.1603G} {497, 1603}

\bibitem[\protect\citeauthoryear{{Griffith}, {Hogg}, {Dalcanton},
  {Hasselquist}, {Ratcliffe}, {Ness}  \& {Weinberg}}{{Griffith}
  et~al.}{2024}]{Griffith2024}
{Griffith} E.~J.,  {Hogg} D.~W.,  {Dalcanton} J.~J.,  {Hasselquist} S.,
  {Ratcliffe} B.,  {Ness} M.,   {Weinberg} D.~H.,  2024, \mn@doi [\aj]
  {10.3847/1538-3881/ad19c7}, \href
  {https://ui.adsabs.harvard.edu/abs/2024AJ....167...98G} {167, 98}

\bibitem[\protect\citeauthoryear{{Gunn} et~al.,}{{Gunn}
  et~al.}{2006}]{Gunn2006}
{Gunn} J.~E.,  et~al., 2006, \mn@doi [\aj] {10.1086/500975}, \href
  {https://ui.adsabs.harvard.edu/abs/2006AJ....131.2332G} {131, 2332}

\bibitem[\protect\citeauthoryear{{Hackshaw}, {Hawkins}, {Filion}, {Horta},
  {Laporte}, {Carr}  \& {Price-Whelan}}{{Hackshaw} et~al.}{2024}]{Hawkshaw2024}
{Hackshaw} Z.,  {Hawkins} K.,  {Filion} C.,  {Horta} D.,  {Laporte} C. F.~P.,
  {Carr} C.,   {Price-Whelan} A.~M.,  2024, \mn@doi [\apj]
  {10.3847/1538-4357/ad900e}, \href
  {https://ui.adsabs.harvard.edu/abs/2024ApJ...977..143H} {977, 143}

\bibitem[\protect\citeauthoryear{{Hawkins}}{{Hawkins}}{2023}]{Hawkins2023}
{Hawkins} K.,  2023, \mn@doi [\mnras] {10.1093/mnras/stad1244}, \href
  {https://ui.adsabs.harvard.edu/abs/2023MNRAS.525.3318H} {525, 3318}

\bibitem[\protect\citeauthoryear{{Hayden} et~al.,}{{Hayden}
  et~al.}{2015}]{Hayden2015}
{Hayden} M.~R.,  et~al., 2015, \mn@doi [\apj] {10.1088/0004-637X/808/2/132},
  \href {http://adsabs.harvard.edu/abs/2015ApJ...808..132H} {808, 132}

\bibitem[\protect\citeauthoryear{{Haywood}}{{Haywood}}{2006}]{Haywood2006}
{Haywood} M.,  2006, \mn@doi [\mnras] {10.1111/j.1365-2966.2006.10802.x}, \href
  {https://ui.adsabs.harvard.edu/abs/2006MNRAS.371.1760H} {371, 1760}

\bibitem[\protect\citeauthoryear{{Haywood}}{{Haywood}}{2008}]{Haywood2008}
{Haywood} M.,  2008, \mn@doi [\mnras] {10.1111/j.1365-2966.2008.13395.x}, \href
  {https://ui.adsabs.harvard.edu/abs/2008MNRAS.388.1175H} {388, 1175}

\bibitem[\protect\citeauthoryear{{Haywood}, {Di Matteo}, {Lehnert}, {Katz}  \&
  {G{\'o}mez}}{{Haywood} et~al.}{2013}]{Haywood2013}
{Haywood} M.,  {Di Matteo} P.,  {Lehnert} M.~D.,  {Katz} D.,   {G{\'o}mez} A.,
  2013, \mn@doi [\aap] {10.1051/0004-6361/201321397}, \href
  {https://ui.adsabs.harvard.edu/abs/2013A&A...560A.109H} {560, A109}

\bibitem[\protect\citeauthoryear{{Haywood}, {Di Matteo}, {Snaith}  \&
  {Lehnert}}{{Haywood} et~al.}{2015}]{Haywood2015}
{Haywood} M.,  {Di Matteo} P.,  {Snaith} O.,   {Lehnert} M.~D.,  2015, \mn@doi
  [\aap] {10.1051/0004-6361/201425459}, \href
  {https://ui.adsabs.harvard.edu/abs/2015A&A...579A...5H} {579, A5}

\bibitem[\protect\citeauthoryear{{Haywood}, {Lehnert}, {Di Matteo}, {Snaith},
  {Schultheis}, {Katz}  \& {G{\'o}mez}}{{Haywood} et~al.}{2016}]{Haywood2016}
{Haywood} M.,  {Lehnert} M.~D.,  {Di Matteo} P.,  {Snaith} O.,  {Schultheis}
  M.,  {Katz} D.,   {G{\'o}mez} A.,  2016, \mn@doi [\aap]
  {10.1051/0004-6361/201527567}, \href
  {https://ui.adsabs.harvard.edu/abs/2016A&A...589A..66H} {589, A66}

\bibitem[\protect\citeauthoryear{{Haywood}, {Di Matteo}, {Lehnert}, {Snaith},
  {Fragkoudi}  \& {Khoperskov}}{{Haywood} et~al.}{2018}]{Haywood2018}
{Haywood} M.,  {Di Matteo} P.,  {Lehnert} M.,  {Snaith} O.,  {Fragkoudi} F.,
  {Khoperskov} S.,  2018, \mn@doi [\aap] {10.1051/0004-6361/201731363}, \href
  {https://ui.adsabs.harvard.edu/abs/2018A&A...618A..78H} {618, A78}

\bibitem[\protect\citeauthoryear{{Horta}, {Price-Whelan}, {Hogg}, {Johnston},
  {Widrow}, {Dalcanton}, {Ness}  \& {Hunt}}{{Horta} et~al.}{2024a}]{Horta2024}
{Horta} D.,  {Price-Whelan} A.~M.,  {Hogg} D.~W.,  {Johnston} K.~V.,  {Widrow}
  L.,  {Dalcanton} J.~J.,  {Ness} M.~K.,   {Hunt} J. A.~S.,  2024a, \mn@doi
  [\apj] {10.3847/1538-4357/ad16e8}, \href
  {https://ui.adsabs.harvard.edu/abs/2024ApJ...962..165H} {962, 165}

\bibitem[\protect\citeauthoryear{{Horta}, {Price-Whelan}, {Hogg}, {Johnston},
  {Widrow}, {Dalcanton}, {Ness}  \& {Hunt}}{{Horta} et~al.}{2024b}]{Horta_oti}
{Horta} D.,  {Price-Whelan} A.~M.,  {Hogg} D.~W.,  {Johnston} K.~V.,  {Widrow}
  L.,  {Dalcanton} J.~J.,  {Ness} M.~K.,   {Hunt} J. A.~S.,  2024b, \mn@doi
  [\apj] {10.3847/1538-4357/ad16e8}, \href
  {https://ui.adsabs.harvard.edu/abs/2024ApJ...962..165H} {962, 165}

\bibitem[\protect\citeauthoryear{{Hunt} \& {Vasiliev}}{{Hunt} \&
  {Vasiliev}}{2025}]{Hunt2025}
{Hunt} J. A.~S.,  {Vasiliev} E.,  2025, \mn@doi [\nar]
  {10.1016/j.newar.2024.101721}, \href
  {https://ui.adsabs.harvard.edu/abs/2025NewAR.10001721H} {100, 101721}

\bibitem[\protect\citeauthoryear{{Imig} et~al.,}{{Imig}
  et~al.}{2023}]{Imig2023}
{Imig} J.,  et~al., 2023, \mn@doi [\apj] {10.3847/1538-4357/ace9b8}, \href
  {https://ui.adsabs.harvard.edu/abs/2023ApJ...954..124I} {954, 124}

\bibitem[\protect\citeauthoryear{{Johnson} et~al.,}{{Johnson}
  et~al.}{2024}]{Johnson2025}
{Johnson} J.~W.,  et~al., 2024, \mn@doi [arXiv e-prints]
  {10.48550/arXiv.2410.13256}, \href
  {https://ui.adsabs.harvard.edu/abs/2024arXiv241013256J} {p. arXiv:2410.13256}

\bibitem[\protect\citeauthoryear{{Karakas} \& {Lattanzio}}{{Karakas} \&
  {Lattanzio}}{2014}]{Karakas2014}
{Karakas} A.~I.,  {Lattanzio} J.~C.,  2014, \mn@doi [\pasa]
  {10.1017/pasa.2014.21}, \href
  {https://ui.adsabs.harvard.edu/abs/2014PASA...31...30K} {31, e030}

\bibitem[\protect\citeauthoryear{{Katz}, {G{\'o}mez}, {Haywood}, {Snaith}  \&
  {Di Matteo}}{{Katz} et~al.}{2021}]{Katz2021}
{Katz} D.,  {G{\'o}mez} A.,  {Haywood} M.,  {Snaith} O.,   {Di Matteo} P.,
  2021, \mn@doi [\aap] {10.1051/0004-6361/202140453}, \href
  {https://ui.adsabs.harvard.edu/abs/2021A&A...655A.111K} {655, A111}

\bibitem[\protect\citeauthoryear{{Khoperskov}, {Haywood}, {Snaith}, {Di
  Matteo}, {Lehnert}, {Vasiliev}, {Naroenkov}  \& {Berczik}}{{Khoperskov}
  et~al.}{2021}]{Kh2021}
{Khoperskov} S.,  {Haywood} M.,  {Snaith} O.,  {Di Matteo} P.,  {Lehnert} M.,
  {Vasiliev} E.,  {Naroenkov} S.,   {Berczik} P.,  2021, \mn@doi [\mnras]
  {10.1093/mnras/staa3996}, \href
  {https://ui.adsabs.harvard.edu/abs/2021MNRAS.501.5176K} {501, 5176}

\bibitem[\protect\citeauthoryear{{Kobayashi}, {Karakas}  \&
  {Lugaro}}{{Kobayashi} et~al.}{2020}]{CK2020}
{Kobayashi} C.,  {Karakas} A.~I.,   {Lugaro} M.,  2020, \mn@doi [\apj]
  {10.3847/1538-4357/abae65}, \href
  {https://ui.adsabs.harvard.edu/abs/2020ApJ...900..179K} {900, 179}

\bibitem[\protect\citeauthoryear{{Kollmeier} et~al.,}{{Kollmeier}
  et~al.}{2025}]{Kollmeier2025}
{Kollmeier} J.~A.,  et~al., 2025, \mn@doi [arXiv e-prints]
  {10.48550/arXiv.2507.06989}, \href
  {https://ui.adsabs.harvard.edu/abs/2025arXiv250706989K} {p. arXiv:2507.06989}

\bibitem[\protect\citeauthoryear{{Kuhn}, {Guo}, {Martin}, {Bayless}, {Gates}
  \& {Puleo}}{{Kuhn} et~al.}{2024}]{Kuhn2024}
{Kuhn} V.,  {Guo} Y.,  {Martin} A.,  {Bayless} J.,  {Gates} E.,   {Puleo} A.,
  2024, \mn@doi [\apjl] {10.3847/2041-8213/ad43eb}, \href
  {https://ui.adsabs.harvard.edu/abs/2024ApJ...968L..15K} {968, L15}

\bibitem[\protect\citeauthoryear{{Laporte}, {Johnston}, {G{\'o}mez},
  {Garavito-Camargo}  \& {Besla}}{{Laporte} et~al.}{2018}]{Laporte2018}
{Laporte} C. F.~P.,  {Johnston} K.~V.,  {G{\'o}mez} F.~A.,  {Garavito-Camargo}
  N.,   {Besla} G.,  2018, \mn@doi [\mnras] {10.1093/mnras/sty1574}, \href
  {https://ui.adsabs.harvard.edu/abs/2018MNRAS.481..286L} {481, 286}

\bibitem[\protect\citeauthoryear{{Lian} et~al.,}{{Lian}
  et~al.}{2020a}]{Lian2020a}
{Lian} J.,  et~al., 2020a, \mn@doi [\mnras] {10.1093/mnras/staa867}, \href
  {https://ui.adsabs.harvard.edu/abs/2020MNRAS.494.2561L} {494, 2561}

\bibitem[\protect\citeauthoryear{{Lian} et~al.,}{{Lian}
  et~al.}{2020b}]{Lian2020}
{Lian} J.,  et~al., 2020b, \mn@doi [\mnras] {10.1093/mnras/staa2078}, \href
  {https://ui.adsabs.harvard.edu/abs/2020MNRAS.497.2371L} {497, 2371}

\bibitem[\protect\citeauthoryear{{Lian} et~al.,}{{Lian}
  et~al.}{2022}]{Lian2022}
{Lian} J.,  et~al., 2022, \mn@doi [\mnras] {10.1093/mnras/stac1151}, \href
  {https://ui.adsabs.harvard.edu/abs/2022MNRAS.513.4130L} {513, 4130}

\bibitem[\protect\citeauthoryear{{Lu}, {Ness}, {Buck}, {Zinn}  \&
  {Johnston}}{{Lu} et~al.}{2022a}]{Lu2022a}
{Lu} Y.~L.,  {Ness} M.~K.,  {Buck} T.,  {Zinn} J.~C.,   {Johnston} K.~V.,
  2022a, \mn@doi [\mnras] {10.1093/mnras/stac610}, \href
  {https://ui.adsabs.harvard.edu/abs/2022MNRAS.512.2890L} {512, 2890}

\bibitem[\protect\citeauthoryear{{Lu}, {Ness}, {Buck}  \& {Carr}}{{Lu}
  et~al.}{2022b}]{Lu2022}
{Lu} Y.~L.,  {Ness} M.~K.,  {Buck} T.,   {Carr} C.,  2022b, \mn@doi [\mnras]
  {10.1093/mnras/stac780}, \href
  {https://ui.adsabs.harvard.edu/abs/2022MNRAS.512.4697L} {512, 4697}

\bibitem[\protect\citeauthoryear{{Lu} et~al.,}{{Lu} et~al.}{2024}]{Lu2024}
{Lu} Y.~L.,  et~al., 2024, \mn@doi [\mnras] {10.1093/mnras/stae2364}, \href
  {https://ui.adsabs.harvard.edu/abs/2024MNRAS.535..392L} {535, 392}

\bibitem[\protect\citeauthoryear{{Mackereth} et~al.,}{{Mackereth}
  et~al.}{2017}]{Mackereth2017}
{Mackereth} J.~T.,  et~al., 2017, \mn@doi [\mnras] {10.1093/mnras/stx1774},
  \href {https://ui.adsabs.harvard.edu/abs/2017MNRAS.471.3057M} {471, 3057}

\bibitem[\protect\citeauthoryear{{Mead}, {De La Garza}  \& {Ness}}{{Mead}
  et~al.}{2025}]{Mead2025}
{Mead} J.,  {De La Garza} R.,   {Ness} M.,  2025, \mn@doi [arXiv e-prints]
  {10.48550/arXiv.2504.18532}, \href
  {https://ui.adsabs.harvard.edu/abs/2025arXiv250418532M} {p. arXiv:2504.18532}

\bibitem[\protect\citeauthoryear{{Miglio} et~al.,}{{Miglio}
  et~al.}{2021}]{Miglio2021}
{Miglio} A.,  et~al., 2021, \mn@doi [\aap] {10.1051/0004-6361/202038307}, \href
  {https://ui.adsabs.harvard.edu/abs/2021A&A...645A..85M} {645, A85}

\bibitem[\protect\citeauthoryear{{Minchev}, {Famaey}, {Combes}, {Di Matteo},
  {Mouhcine}  \& {Wozniak}}{{Minchev} et~al.}{2011}]{Minchev2011}
{Minchev} I.,  {Famaey} B.,  {Combes} F.,  {Di Matteo} P.,  {Mouhcine} M.,
  {Wozniak} H.,  2011, \mn@doi [\aap] {10.1051/0004-6361/201015139}, \href
  {http://adsabs.harvard.edu/abs/2011A%26A...527A.147M} {527, A147}

\bibitem[\protect\citeauthoryear{{Minchev}, {Chiappini}  \& {Martig}}{{Minchev}
  et~al.}{2013}]{Minchev2013}
{Minchev} I.,  {Chiappini} C.,   {Martig} M.,  2013, \mn@doi [\aap]
  {10.1051/0004-6361/201220189}, \href
  {https://ui.adsabs.harvard.edu/abs/2013A&A...558A...9M} {558, A9}

\bibitem[\protect\citeauthoryear{{Minchev}, {Steinmetz}, {Chiappini}, {Martig},
  {Anders}, {Matijevic}  \& {de Jong}}{{Minchev} et~al.}{2017}]{Minchev2017}
{Minchev} I.,  {Steinmetz} M.,  {Chiappini} C.,  {Martig} M.,  {Anders} F.,
  {Matijevic} G.,   {de Jong} R.~S.,  2017, \mn@doi [\apj]
  {10.3847/1538-4357/834/1/27}, \href
  {https://ui.adsabs.harvard.edu/abs/2017ApJ...834...27M} {834, 27}

\bibitem[\protect\citeauthoryear{{Minchev} et~al.,}{{Minchev}
  et~al.}{2018}]{Minchev2018}
{Minchev} I.,  et~al., 2018, \mn@doi [\mnras] {10.1093/mnras/sty2033}, \href
  {https://ui.adsabs.harvard.edu/abs/2018MNRAS.481.1645M} {481, 1645}

\bibitem[\protect\citeauthoryear{{Ness}, {Hogg}, {Rix}, {Martig},
  {Pinsonneault}  \& {Ho}}{{Ness} et~al.}{2016}]{Ness2016}
{Ness} M.,  {Hogg} D.~W.,  {Rix} H.-W.,  {Martig} M.,  {Pinsonneault} M.~H.,
  {Ho} A.~Y.~Q.,  2016, \mn@doi [\apj] {10.3847/0004-637X/823/2/114}, \href
  {http://adsabs.harvard.edu/abs/2016ApJ...823..114N} {823, 114}

\bibitem[\protect\citeauthoryear{{Ness}, {Johnston}, {Blancato}, {Rix},
  {Beane}, {Bird}  \& {Hawkins}}{{Ness} et~al.}{2019}]{Ness2019}
{Ness} M.~K.,  {Johnston} K.~V.,  {Blancato} K.,  {Rix} H.~W.,  {Beane} A.,
  {Bird} J.~C.,   {Hawkins} K.,  2019, \mn@doi [\apj]
  {10.3847/1538-4357/ab3e3c}, \href
  {https://ui.adsabs.harvard.edu/abs/2019ApJ...883..177N} {883, 177}

\bibitem[\protect\citeauthoryear{{Ness}, {Wheeler}, {McKinnon}, {Horta},
  {Casey}, {Cunningham}  \& {Price-Whelan}}{{Ness} et~al.}{2022}]{Ness2022}
{Ness} M.~K.,  {Wheeler} A.~J.,  {McKinnon} K.,  {Horta} D.,  {Casey} A.~R.,
  {Cunningham} E.~C.,   {Price-Whelan} A.~M.,  2022, \mn@doi [\apj]
  {10.3847/1538-4357/ac4754}, \href
  {https://ui.adsabs.harvard.edu/abs/2022ApJ...926..144N} {926, 144}

\bibitem[\protect\citeauthoryear{{Ness} et~al.,}{{Ness}
  et~al.}{2026}]{Ness2026}
{Ness} M.~K.,  et~al., 2026, \mn@doi [\mnras] {10.1093/mnras/stag637}, \href
  {https://ui.adsabs.harvard.edu/abs/2026MNRAS.tmp..626N} {}

\bibitem[\protect\citeauthoryear{{Nidever} et~al.,}{{Nidever}
  et~al.}{2014}]{Nidever2014}
{Nidever} D.~L.,  et~al., 2014, \mn@doi [\apj] {10.1088/0004-637X/796/1/38},
  \href {http://adsabs.harvard.edu/abs/2014ApJ...796...38N} {796, 38}

\bibitem[\protect\citeauthoryear{{Nomoto}, {Iwamoto}, {Nakasato}, {Thielemann},
  {Brachwitz}, {Tsujimoto}, {Kubo}  \& {Kishimoto}}{{Nomoto}
  et~al.}{1997}]{Nomoto1997}
{Nomoto} K.,  {Iwamoto} K.,  {Nakasato} N.,  {Thielemann} F.~K.,  {Brachwitz}
  F.,  {Tsujimoto} T.,  {Kubo} Y.,   {Kishimoto} N.,  1997, \mn@doi [\nphysa]
  {10.1016/S0375-9474(97)00291-1}, \href
  {https://ui.adsabs.harvard.edu/abs/1997NuPhA.621..467N} {621, 467}

\bibitem[\protect\citeauthoryear{{Nomoto}, {Tominaga}, {Umeda}, {Kobayashi}  \&
  {Maeda}}{{Nomoto} et~al.}{2006}]{Nomoto2006}
{Nomoto} K.,  {Tominaga} N.,  {Umeda} H.,  {Kobayashi} C.,   {Maeda} K.,  2006,
  \mn@doi [\nphysa] {10.1016/j.nuclphysa.2006.05.008}, \href
  {https://ui.adsabs.harvard.edu/abs/2006NuPhA.777..424N} {777, 424}

\bibitem[\protect\citeauthoryear{{Nomoto}, {Kobayashi}  \& {Tominaga}}{{Nomoto}
  et~al.}{2013}]{Nomoto2013}
{Nomoto} K.,  {Kobayashi} C.,   {Tominaga} N.,  2013, \mn@doi [\araa]
  {10.1146/annurev-astro-082812-140956}, \href
  {https://ui.adsabs.harvard.edu/abs/2013ARA&A..51..457N} {51, 457}

\bibitem[\protect\citeauthoryear{{Orkney}, {Laporte}, {Grand}  \&
  {Springel}}{{Orkney} et~al.}{2025a}]{Orkney2025}
{Orkney} M. D.~A.,  {Laporte} C. F.~P.,  {Grand} R. J.~J.,   {Springel} V.,
  2025a, \mn@doi [arXiv e-prints] {10.48550/arXiv.2506.07038}, \href
  {https://ui.adsabs.harvard.edu/abs/2025arXiv250607038O} {p. arXiv:2506.07038}

\bibitem[\protect\citeauthoryear{{Orkney}, {Laporte}, {Grand}  \&
  {Springel}}{{Orkney} et~al.}{2025b}]{Ork2025}
{Orkney} M. D.~A.,  {Laporte} C. F.~P.,  {Grand} R. J.~J.,   {Springel} V.,
  2025b, \mn@doi [arXiv e-prints] {10.48550/arXiv.2506.07038}, \href
  {https://ui.adsabs.harvard.edu/abs/2025arXiv250607038O} {p. arXiv:2506.07038}

\bibitem[\protect\citeauthoryear{{Ortigoza-Urdaneta}
  et~al.,}{{Ortigoza-Urdaneta} et~al.}{2023}]{Orti2023}
{Ortigoza-Urdaneta} M.,  et~al., 2023, \mn@doi [\aap]
  {10.1051/0004-6361/202346325}, \href
  {https://ui.adsabs.harvard.edu/abs/2023A&A...676A.140O} {676, A140}

\bibitem[\protect\citeauthoryear{{Parul}, {Bailin}, {Loebman}, {Wetzel},
  {Barry}  \& {Bhattarai}}{{Parul} et~al.}{2025}]{Hanna2025}
{Parul} H.,  {Bailin} J.,  {Loebman} S.~R.,  {Wetzel} A.,  {Barry} M.,
  {Bhattarai} B.,  2025, \mn@doi [\mnras] {10.1093/mnras/staf137}, \href
  {https://ui.adsabs.harvard.edu/abs/2025MNRAS.537.1571P} {537, 1571}

\bibitem[\protect\citeauthoryear{{Queiroz} et~al.,}{{Queiroz}
  et~al.}{2020}]{Queiroz2020}
{Queiroz} A.~B.~A.,  et~al., 2020, \mn@doi [\aap]
  {10.1051/0004-6361/201937364}, \href
  {https://ui.adsabs.harvard.edu/abs/2020A&A...638A..76Q} {638, A76}

\bibitem[\protect\citeauthoryear{{Queiroz} et~al.,}{{Queiroz}
  et~al.}{2023}]{Queiroz2023}
{Queiroz} A.~B.~A.,  et~al., 2023, \mn@doi [\aap]
  {10.1051/0004-6361/202245399}, \href
  {https://ui.adsabs.harvard.edu/abs/2023A&A...673A.155Q} {673, A155}

\bibitem[\protect\citeauthoryear{{Ratcliffe}, {Ness}, {Buck}, {Johnston},
  {Sen}, {Beraldo e Silva}  \& {Debattista}}{{Ratcliffe}
  et~al.}{2021}]{Ratcliffe2021}
{Ratcliffe} B.~L.,  {Ness} M.~K.,  {Buck} T.,  {Johnston} K.~V.,  {Sen} B.,
  {Beraldo e Silva} L.,   {Debattista} V.~P.,  2021, arXiv e-prints, \href
  {https://ui.adsabs.harvard.edu/abs/2021arXiv210708088R} {p. arXiv:2107.08088}

\bibitem[\protect\citeauthoryear{{Ratcliffe} et~al.,}{{Ratcliffe}
  et~al.}{2023}]{Ratcliffe2023}
{Ratcliffe} B.,  et~al., 2023, \mn@doi [\mnras] {10.1093/mnras/stad1573}, \href
  {https://ui.adsabs.harvard.edu/abs/2023MNRAS.525.2208R} {525, 2208}

\bibitem[\protect\citeauthoryear{{Ratcliffe}, {Minchev}, {Cescutti}, {Spitoni},
  {J{\"o}nsson}, {Anders}, {Queiroz}  \& {Steinmetz}}{{Ratcliffe}
  et~al.}{2024}]{Ratcliffe2024}
{Ratcliffe} B.,  {Minchev} I.,  {Cescutti} G.,  {Spitoni} E.,  {J{\"o}nsson}
  H.,  {Anders} F.,  {Queiroz} A.,   {Steinmetz} M.,  2024, \mn@doi [\mnras]
  {10.1093/mnras/stae226}, \href
  {https://ui.adsabs.harvard.edu/abs/2024MNRAS.528.3464R} {528, 3464}

\bibitem[\protect\citeauthoryear{{Rix} \& {Bovy}}{{Rix} \&
  {Bovy}}{2013}]{Rix2013}
{Rix} H.-W.,  {Bovy} J.,  2013, \mn@doi [\aapr] {10.1007/s00159-013-0061-8},
  \href {https://ui.adsabs.harvard.edu/abs/2013A&ARv..21...61R} {21, 61}

\bibitem[\protect\citeauthoryear{{Ro{\v s}kar}, {Debattista}, {Quinn},
  {Stinson}  \& {Wadsley}}{{Ro{\v s}kar} et~al.}{2008}]{Roskar2008}
{Ro{\v s}kar} R.,  {Debattista} V.~P.,  {Quinn} T.~R.,  {Stinson} G.~S.,
  {Wadsley} J.,  2008, \mn@doi [\apjl] {10.1086/592231}, \href
  {http://adsabs.harvard.edu/abs/2008ApJ...684L..79R} {684, L79}

\bibitem[\protect\citeauthoryear{{Ruiz-Lara}, {Gallart}, {Bernard}  \&
  {Cassisi}}{{Ruiz-Lara} et~al.}{2020}]{RL2020}
{Ruiz-Lara} T.,  {Gallart} C.,  {Bernard} E.~J.,   {Cassisi} S.,  2020, \mn@doi
  [Nature Astronomy] {10.1038/s41550-020-1097-0}, \href
  {https://ui.adsabs.harvard.edu/abs/2020NatAs...4..965R} {4, 965}

\bibitem[\protect\citeauthoryear{{SDSS Collaboration} et~al.,}{{SDSS
  Collaboration} et~al.}{2025}]{DR19}
{SDSS Collaboration} et~al., 2025, \mn@doi [arXiv e-prints]
  {10.48550/arXiv.2507.07093}, \href
  {https://ui.adsabs.harvard.edu/abs/2025arXiv250707093S} {p. arXiv:2507.07093}

\bibitem[\protect\citeauthoryear{{S{\'a}nchez-Menguiano}, {S{\'a}nchez},
  {S{\'a}nchez Almeida}  \& {Mu{\~n}oz-Tu{\~n}{\'o}n}}{{S{\'a}nchez-Menguiano}
  et~al.}{2024}]{LSM2024}
{S{\'a}nchez-Menguiano} L.,  {S{\'a}nchez} S.~F.,  {S{\'a}nchez Almeida} J.,
  {Mu{\~n}oz-Tu{\~n}{\'o}n} C.,  2024, \mn@doi [\aap]
  {10.1051/0004-6361/202348423}, \href
  {https://ui.adsabs.harvard.edu/abs/2024A&A...682L..11S} {682, L11}

\bibitem[\protect\citeauthoryear{{Sanders} \& {Das}}{{Sanders} \&
  {Das}}{2018}]{SandersDas2018}
{Sanders} J.~L.,  {Das} P.,  2018, \mn@doi [\mnras] {10.1093/mnras/sty2490},
  \href {https://ui.adsabs.harvard.edu/abs/2018MNRAS.481.4093S} {481, 4093}

\bibitem[\protect\citeauthoryear{{Sch{\"o}nrich} \& {Binney}}{{Sch{\"o}nrich}
  \& {Binney}}{2009a}]{RS2009}
{Sch{\"o}nrich} R.,  {Binney} J.,  2009a, \mn@doi [\mnras]
  {10.1111/j.1365-2966.2009.14750.x}, \href
  {https://ui.adsabs.harvard.edu/abs/2009MNRAS.396..203S} {396, 203}

\bibitem[\protect\citeauthoryear{{Sch{\"o}nrich} \& {Binney}}{{Sch{\"o}nrich}
  \& {Binney}}{2009b}]{Schonrich2009}
{Sch{\"o}nrich} R.,  {Binney} J.,  2009b, \mn@doi [\mnras]
  {10.1111/j.1365-2966.2009.14750.x}, \href
  {https://ui.adsabs.harvard.edu/abs/2009MNRAS.396..203S} {396, 203}

\bibitem[\protect\citeauthoryear{{Sch{\"o}nrich} \& {McMillan}}{{Sch{\"o}nrich}
  \& {McMillan}}{2017}]{Schonrich2017}
{Sch{\"o}nrich} R.,  {McMillan} P.~J.,  2017, \mn@doi [\mnras]
  {10.1093/mnras/stx093}, \href
  {https://ui.adsabs.harvard.edu/abs/2017MNRAS.467.1154S} {467, 1154}

\bibitem[\protect\citeauthoryear{{Sch{\"o}nrich}, {Binney}  \&
  {Dehnen}}{{Sch{\"o}nrich} et~al.}{2010}]{Schonrich2010}
{Sch{\"o}nrich} R.,  {Binney} J.,   {Dehnen} W.,  2010, \mn@doi [\mnras]
  {10.1111/j.1365-2966.2010.16253.x}, \href
  {https://ui.adsabs.harvard.edu/abs/2010MNRAS.403.1829S} {403, 1829}

\bibitem[\protect\citeauthoryear{{Sellwood} \& {Binney}}{{Sellwood} \&
  {Binney}}{2002}]{Selwood2002}
{Sellwood} J.~A.,  {Binney} J.~J.,  2002, \mn@doi [\mnras]
  {10.1046/j.1365-8711.2002.05806.x}, \href
  {http://adsabs.harvard.edu/abs/2002MNRAS.336..785S} {336, 785}

\bibitem[\protect\citeauthoryear{{Sharma}, {Hayden}  \&
  {Bland-Hawthorn}}{{Sharma} et~al.}{2021}]{Sharma2021}
{Sharma} S.,  {Hayden} M.~R.,   {Bland-Hawthorn} J.,  2021, \mn@doi [\mnras]
  {10.1093/mnras/stab2015}, \href
  {https://ui.adsabs.harvard.edu/abs/2021MNRAS.507.5882S} {507, 5882}

\bibitem[\protect\citeauthoryear{{Sharma} et~al.,}{{Sharma}
  et~al.}{2022}]{Sharma2022}
{Sharma} S.,  et~al., 2022, \mn@doi [\mnras] {10.1093/mnras/stab3341}, \href
  {https://ui.adsabs.harvard.edu/abs/2022MNRAS.510..734S} {510, 734}

\bibitem[\protect\citeauthoryear{{Silva Aguirre} et~al.,}{{Silva Aguirre}
  et~al.}{2018}]{SilvaAguirre2018}
{Silva Aguirre} V.,  et~al., 2018, \mn@doi [\mnras] {10.1093/mnras/sty150},
  \href {https://ui.adsabs.harvard.edu/abs/2018MNRAS.475.5487S} {475, 5487}

\bibitem[\protect\citeauthoryear{{Snaith}, {Haywood}, {Di Matteo}, {Lehnert},
  {Combes}, {Katz}  \& {G{\'o}mez}}{{Snaith} et~al.}{2015}]{Snaith2015}
{Snaith} O.,  {Haywood} M.,  {Di Matteo} P.,  {Lehnert} M.~D.,  {Combes} F.,
  {Katz} D.,   {G{\'o}mez} A.,  2015, \mn@doi [\aap]
  {10.1051/0004-6361/201424281}, \href
  {https://ui.adsabs.harvard.edu/abs/2015A&A...578A..87S} {578, A87}

\bibitem[\protect\citeauthoryear{{Snaith}, {Haywood}, {Di Matteo}, {Lehnert},
  {Katz}  \& {Khoperskov}}{{Snaith} et~al.}{2022}]{Snaith2022}
{Snaith} O.,  {Haywood} M.,  {Di Matteo} P.,  {Lehnert} M.,  {Katz} D.,
  {Khoperskov} S.,  2022, \mn@doi [\aap] {10.1051/0004-6361/202039526}, \href
  {https://ui.adsabs.harvard.edu/abs/2022A&A...659A..64S} {659, A64}

\bibitem[\protect\citeauthoryear{{Spitoni}, {Silva Aguirre}, {Matteucci},
  {Calura}  \& {Grisoni}}{{Spitoni} et~al.}{2019}]{Spitoni2019}
{Spitoni} E.,  {Silva Aguirre} V.,  {Matteucci} F.,  {Calura} F.,   {Grisoni}
  V.,  2019, \mn@doi [\aap] {10.1051/0004-6361/201834188}, \href
  {https://ui.adsabs.harvard.edu/abs/2019A&A...623A..60S} {623, A60}

\bibitem[\protect\citeauthoryear{{Spitoni} et~al.,}{{Spitoni}
  et~al.}{2021}]{Spitoni2021}
{Spitoni} E.,  et~al., 2021, \mn@doi [\aap] {10.1051/0004-6361/202039864},
  \href {https://ui.adsabs.harvard.edu/abs/2021A&A...647A..73S} {647, A73}

\bibitem[\protect\citeauthoryear{{Spitoni}, {Aguirre B{\o}rsen-Koch}, {Verma}
  \& {Stokholm}}{{Spitoni} et~al.}{2022}]{Spitoni2022}
{Spitoni} E.,  {Aguirre B{\o}rsen-Koch} V.,  {Verma} K.,   {Stokholm} A.,
  2022, \mn@doi [\aap] {10.1051/0004-6361/202142469}, \href
  {https://ui.adsabs.harvard.edu/abs/2022A&A...663A.174S} {663, A174}

\bibitem[\protect\citeauthoryear{{Spitoni} et~al.,}{{Spitoni}
  et~al.}{2023}]{Spitoni2023}
{Spitoni} E.,  et~al., 2023, \mn@doi [\aap] {10.1051/0004-6361/202244349},
  \href {https://ui.adsabs.harvard.edu/abs/2023A&A...670A.109S} {670, A109}

\bibitem[\protect\citeauthoryear{{Spitoni}, {Matteucci}, {Gratton},
  {Ratcliffe}, {Minchev}  \& {Cescutti}}{{Spitoni} et~al.}{2024}]{ES2024}
{Spitoni} E.,  {Matteucci} F.,  {Gratton} R.,  {Ratcliffe} B.,  {Minchev} I.,
  {Cescutti} G.,  2024, \mn@doi [\aap] {10.1051/0004-6361/202450754}, \href
  {https://ui.adsabs.harvard.edu/abs/2024A&A...690A.208S} {690, A208}

\bibitem[\protect\citeauthoryear{{Stone-Martinez}, {Holtzman}, {Yuxi}, {Lu},
  {Imig}, {Griffith}, {Bellinger}  \& {Saydjari}}{{Stone-Martinez}
  et~al.}{2025}]{StoneM2025}
{Stone-Martinez} A.,  {Holtzman} J.~A.,  {Yuxi} {Lu} {Imig} S. H.~J.,
  {Griffith} E.~J.,  {Bellinger} E.,   {Saydjari} A.~K.,  2025, \mn@doi [arXiv
  e-prints] {10.48550/arXiv.2503.03138}, \href
  {https://ui.adsabs.harvard.edu/abs/2025arXiv250303138S} {p. arXiv:2503.03138}

\bibitem[\protect\citeauthoryear{{Tolstoy}, {Hill}  \& {Tosi}}{{Tolstoy}
  et~al.}{2009}]{Tolstoy2009}
{Tolstoy} E.,  {Hill} V.,   {Tosi} M.,  2009, \mn@doi [\araa]
  {10.1146/annurev-astro-082708-101650}, \href
  {https://ui.adsabs.harvard.edu/abs/2009ARA&A..47..371T} {47, 371}

\bibitem[\protect\citeauthoryear{{Tsukui}, {Wisnioski}, {Bland-Hawthorn}  \&
  {Freeman}}{{Tsukui} et~al.}{2025}]{Takafumi2025}
{Tsukui} T.,  {Wisnioski} E.,  {Bland-Hawthorn} J.,   {Freeman} K.,  2025,
  \mn@doi [\mnras] {10.1093/mnras/staf604}, \href
  {https://ui.adsabs.harvard.edu/abs/2025MNRAS.540.3493T} {540, 3493}

\bibitem[\protect\citeauthoryear{{Villalobos}, {Kazantzidis}  \&
  {Helmi}}{{Villalobos} et~al.}{2010}]{Villa2010}
{Villalobos} {\'A}.,  {Kazantzidis} S.,   {Helmi} A.,  2010, \mn@doi [\apj]
  {10.1088/0004-637X/718/1/314}, \href
  {https://ui.adsabs.harvard.edu/abs/2010ApJ...718..314V} {718, 314}

\bibitem[\protect\citeauthoryear{{Vincenzo} \& {Kobayashi}}{{Vincenzo} \&
  {Kobayashi}}{2020}]{Vincenzo2020}
{Vincenzo} F.,  {Kobayashi} C.,  2020, \mn@doi [\mnras]
  {10.1093/mnras/staa1451}, \href
  {https://ui.adsabs.harvard.edu/abs/2020MNRAS.496...80V} {496, 80}

\bibitem[\protect\citeauthoryear{{Wang}, {Carrillo}, {Ness}  \& {Buck}}{{Wang}
  et~al.}{2024}]{Wang2024}
{Wang} K.,  {Carrillo} A.,  {Ness} M.~K.,   {Buck} T.,  2024, \mn@doi [\mnras]
  {10.1093/mnras/stad3182}, \href
  {https://ui.adsabs.harvard.edu/abs/2024MNRAS.527..321W} {527, 321}

\bibitem[\protect\citeauthoryear{{Weinberg} et~al.,}{{Weinberg}
  et~al.}{2019}]{Weinberg2019}
{Weinberg} D.~H.,  et~al., 2019, \mn@doi [\apj] {10.3847/1538-4357/ab07c7},
  \href {https://ui.adsabs.harvard.edu/abs/2019ApJ...874..102W} {874, 102}

\bibitem[\protect\citeauthoryear{{Wheeler}, {Abril-Cabezas}, {Trick},
  {Fragkoudi}  \& {Ness}}{{Wheeler} et~al.}{2022}]{Wheeler2022a}
{Wheeler} A.,  {Abril-Cabezas} I.,  {Trick} W.~H.,  {Fragkoudi} F.,   {Ness}
  M.,  2022, \mn@doi [\apj] {10.3847/1538-4357/ac7da0}, \href
  {https://ui.adsabs.harvard.edu/abs/2022ApJ...935...28W} {935, 28}

\bibitem[\protect\citeauthoryear{{Wilson} et~al.,}{{Wilson}
  et~al.}{2019}]{Wilson2019}
{Wilson} J.~C.,  et~al., 2019, \mn@doi [\pasp] {10.1088/1538-3873/ab0075},
  \href {https://ui.adsabs.harvard.edu/abs/2019PASP..131e5001W} {131, 055001}

\bibitem[\protect\citeauthoryear{{Woosley} \& {Weaver}}{{Woosley} \&
  {Weaver}}{1995}]{Woosley1995}
{Woosley} S.~E.,  {Weaver} T.~A.,  1995, \mn@doi [\apjs] {10.1086/192237},
  \href {https://ui.adsabs.harvard.edu/abs/1995ApJS..101..181W} {101, 181}

\bibitem[\protect\citeauthoryear{{Xiang} \& {Rix}}{{Xiang} \&
  {Rix}}{2022}]{Xiang2022}
{Xiang} M.,  {Rix} H.-W.,  2022, \mn@doi [\nat] {10.1038/s41586-022-04496-5},
  \href {https://ui.adsabs.harvard.edu/abs/2022Natur.603..599X} {603, 599}

\bibitem[\protect\citeauthoryear{{Zhang} et~al.,}{{Zhang}
  et~al.}{2025}]{Zhang2025}
{Zhang} H.,  et~al., 2025, \mn@doi [\apjl] {10.3847/2041-8213/adc261}, \href
  {https://ui.adsabs.harvard.edu/abs/2025ApJ...983L..10Z} {983, L10}

\makeatother
\end{thebibliography}
\end{document}